\documentclass[10pt,conference]{IEEEtran}
\usepackage{algorithm,algorithmic} % ACM推荐算法包
\usepackage[numbers, square, comma, sort&compress]{natbib}
\usepackage{graphicx,xcolor}
\usepackage{tikz}
\usepackage{url}
\usepackage{multirow}
\usepackage{amsmath,booktabs}
\usepackage{enumitem}
\usepackage{subcaption}  
\usepackage[flushleft]{threeparttable}

\usepackage{hyperref}
\usepackage{fancyhdr}
\newfloat{procedure}{t}{lop}  % 定义浮动体
\floatname{procedure}{Procedure}  % 强制名称首字母大写
\definecolor{lightergray}{rgb}{0.85, 0.85, 0.85}

\usepackage{CJKutf8}

\newif\ifdraft
\draftfalse  % 改成 \draftfalse 就能隐藏所有批注

\newcommand{\ZZY}[1]{%
  \ifdraft
    \begin{CJK*}{UTF8}{gbsn}%
      \textcolor{red}{\footnotesize\bfseries [#1]}%
    \end{CJK*}%
  \fi
}
\ifodd 0
  \DeclareRobustCommand{\hn}[1]{\textcolor{blue}{#1}} % 高亮模式
\else
  \DeclareRobustCommand{\hn}[1]{#1}% 关闭高亮
\fi
\begin{document}

\title{VRExplorer: A Model-based Approach for \hn{Semi-Automated} Testing of Virtual Reality Scenes}
% VRExplorer: A Model-based Approach for Semi-Automated Testing of Virtual Reality Scenes
% \author{
%     \IEEEauthorblockN{Anonymous Authors}
%    % \IEEEauthorblockA{Paper under double-blind review}
% }

% \author{
%     \IEEEauthorblockN{
%         Zhengyang Zhu\IEEEauthorrefmark{1}\IEEEauthorrefmark{3},
%         Hong-Ning Dai\IEEEauthorrefmark{2}\thanks{*Corresponding author.},
%         Hanyang Guo\IEEEauthorrefmark{1},\\
%         Zeqin Liao\IEEEauthorrefmark{1},
%         Zibin Zheng\IEEEauthorrefmark{1}
%     }
%     \IEEEauthorblockA{\IEEEauthorrefmark{1}School of Software Engineering, Sun Yat-sen University, Zhuhai, China\\
%     zhuzhy57@mail2.sysu.edu.cn, guohy36@mail2.sysu.edu.cn, liaozq8@mail2.sysu.edu.cn, zhzibin@mail.sysu.edu.cn}

%     \IEEEauthorblockA{\IEEEauthorrefmark{2}Department of Computer Science, Hong Kong Baptist University, Hong Kong, China\\
%     hndai@ieee.org}

%     \IEEEauthorblockA{\IEEEauthorrefmark{3}Peng Cheng Laboratory, Shenzhen, China}
% }

\author{
    \IEEEauthorblockN{
        Zhengyang Zhu\IEEEauthorrefmark{1}\IEEEauthorrefmark{3},
        Hong-Ning Dai\IEEEauthorrefmark{2}\thanks{*Corresponding author.},
        Hanyang Guo\IEEEauthorrefmark{1},
        Zeqin Liao\IEEEauthorrefmark{1},
        Zibin Zheng\IEEEauthorrefmark{1}
    }
    \IEEEauthorblockA{\IEEEauthorrefmark{1}\textit{School of Software Engineering, Sun Yat-sen University}, Zhuhai, China\\
    \{zhuzhy57, guohy36, liaozq8\}@mail2.sysu.edu.cn, zhzibin@mail.sysu.edu.cn}

    \IEEEauthorblockA{\IEEEauthorrefmark{2}\textit{Department of Computer Science, Hong Kong Baptist University}, Hong Kong SAR, hndai@ieee.org}

    \IEEEauthorblockA{\IEEEauthorrefmark{3}\textit{Peng Cheng Laboratory}, Shenzhen, China}
}
\fancypagestyle{arxivpreprint}{%
  \fancyhf{}%
  \fancyhead[C]{\scriptsize 2025 40th IEEE/ACM International Conference on Automated Software Engineering (ASE)}%
  \fancyfoot[L]{\scriptsize\textsc{Author Accepted Manuscript}}%
  \fancyfoot[C]{\scriptsize\href{https://doi.org/10.1109/ASE63991.2025.00047}{DOI: 10.1109/ASE63991.2025.00047}}%
  \fancyfoot[R]{\scriptsize\thepage}%
  \renewcommand{\headrulewidth}{0.2pt}%
  \renewcommand{\footrulewidth}{0.2pt}%
}

\maketitle
\pagestyle{arxivpreprint}
\thispagestyle{arxivpreprint}

\begin{abstract}
%Virtual Reality (VR) technologies have received increasing attention recently. 
With the proliferation of Virtual Reality (VR) markets, VR applications are rapidly expanding in scale and complexity, thereby driving an urgent need for assuring VR software quality. Different from traditional mobile applications and computer software, VR testing faces unique challenges due to diverse interactions with virtual objects, complex 3D virtual environments, and intricate sequences to complete tasks. All of these emerging challenges hinder existing VR testing tools from effectively and systematically testing VR applications.
In this paper, we present \textit{VRExplorer}, a novel model-based testing tool to effectively interact with diverse virtual objects and explore complex VR scenes.
Particularly, we design the \underline{E}ntity, \underline{A}ction, and \underline{T}ask (EAT) framework for modeling diverse VR interactions in a generic way. 
Built upon the EAT framework, we then present the \textit{VRExplorer} agent, which can achieve effective scene exploration by incorporating meticulously designed path-finding algorithms into Unity's NavMesh. Moreover, the \textit{VRExplorer} agent can also systematically execute interaction decisions on top of the Probabilistic Finite State Machine (PFSM). 
Experimental evaluation on \hn{11 representative} VR projects shows that \textit{VRExplorer} consistently outperforms the state-of-the-art (SOTA) approach \textit{VRGuide} by achieving significantly higher coverage and better efficiency. Specifically, \textit{VRExplorer} yields up to \hn{122.8\% and 52.8\%} improvements over \textit{VRGuide} in terms of executable lines of code (ELOC) coverage and method (function) coverage, respectively. 
Furthermore, ablation results also verify the essential contributions of each designed module.
More importantly, our \textit{VRExplorer} has successfully detected two \textit{functional} bugs and one \textit{non-functional} bug from real-world projects.
\end{abstract}

\begin{IEEEkeywords}
Software Testing, Virtual Reality, Model-based Testing, Scene Exploration
\end{IEEEkeywords}

% IEEE requires this notice to be visible on the first screen of the posted
% accepted manuscript. Place it between the abstract and the main text.
\par\smallskip
\noindent\begin{minipage}{\columnwidth}
\scriptsize
\copyright\ 2025 IEEE. Personal use of this material is permitted. Permission
from IEEE must be obtained for all other uses, in any current or future media,
including reprinting/republishing this material for advertising or promotional
purposes, creating new collective works, for resale or redistribution to
servers or lists, or reuse of any copyrighted component of this work in other
works.
\end{minipage}
\par\smallskip

\section{Introduction}

% VR定义、市场增长
Virtual reality (VR), working together with other relevant technologies, such as Augmented Reality (AR) and Extended Reality (XR), aims to provide users with an immersive mixed reality (MR) experience~\cite{Overview_On_Hardware_Characteristics_Of_Virtual_Reality_Systems}. 
%refers to computer-based technology that simulates visual, auditory, and tactile effects of an artificially created environment, with the aim of immersing the user in a perceived reality.
Current VR applications have proliferated in diverse fields, such as medical treatment, education, audiovisual entertainment, training simulations, manufacturing, and gaming~\cite{AirwayVR_Virtual_Reality_Trainer_for_Endotracheal_Intubation, Immersive_virtual_reality_as_a_pedagogical_tool_in_education_a_systematic_literature_review_of_quantitative_learning_outcomes_and_experimental_design, Virtual_reality_system_for_industrial_training, WANG2024777, Overview_On_Hardware_Characteristics_Of_Virtual_Reality_Systems, Virtual_reality_and_the_CAVE_Taxonomy_interaction_challenges_and_research_directions}. According to Fortune's market report~\cite{fortune2024vrmarket}, the global VR market is projected to grow from \$44.4 billion in 2025 to \$244.84 billion in 2032. 

%Therefore, ensuring VR application quality has become critically important. particularly those deployed on standalone devices such as the Meta Quest 2~\cite{MetaQuest2}

% 动机：VR Bug有和移动应用类似的问题，并且也有许多是由Unity开发过程中引起的。VR Bug检测是非常费时费力的，自动化VR测试工具来探索场景、交互物体是很有意义的
%VR devices typically operate on operating systems based on Android and Linux, which may introduce a range of common bugs inherited from Android applications~\cite{ICSE_24_An_Empirical_Study_on_Oculus_Virtual}.

With the proliferation of VR markets, VR applications are also rapidly expanding in terms of scale and complexity. Ensuring the software quality of such complex VR applications becomes an urgent need. 
%In particular, Unity, as the primary engine for VR development, is widely adopted~\cite{Unity_in_VR_appsnado_2023, Unity_in_VR_swiftludus_2023}; however, developers frequently encounter various functional bugs during the development life cycle. These bugs often arise due to the complexities of simulating physical interactions and integrating functionalities through Unity and third-party VR interaction packages.
As a critical procedure of software development, testing can thoroughly evaluate a software to check whether both requirements and functional needs are fulfilled without defects. To cater to this growing demand in VR applications, extensive efforts have been made in VR testing, while many of them still largely rely on manual testing, which remains highly labor-intensive and inefficient~\cite{Virtual_Reality_VR_Automated_Testing_in_the_Wild_A_Case_Study_on_Unity_Based_VR_Applications,PredART}. Most recently, several studies have aimed to test VR applications automatically. As an early attempt, \textit{VRTest}~\cite{VRTest} explores VR scenes by controlling the orientation of the camera to interact with virtual objects. \textit{VRGuide}~\cite{VRGuide} has further improved \textit{VRTest} by optimizing exploration routes to circumvent occluded objects. Besides \textit{VRTest} and \textit{VRGuide}, other researchers also explore using other techniques, such as computer vision (CV) and generative artificial intelligence (GenAI) to achieve scene exploration~\cite{Utilizing_Generative_AI_for_VR_Exploration_Testing_A_Case_Study} or understanding~\cite{li2024groundedguiunderstandingvision}. However, these testing tools are still struggling to test VR applications with increased complexity. 
%However, these testing tools cannot tackle the above three fundamental challenges properly. 

The challenges in VR testing stem from the \textit{unique features} of VR applications, sharply different from conventional mobile applications and computer software. 
First, enabled by diverse peripheral devices (e.g., VR headsets, controllers, joysticks, wands, and haptic gloves), VR systems can support a larger \textit{diversity of interactions} (such as grabbing, pressing, touching, pulling, climbing, and shooting) than conventional mobile systems and PCs~\hn{\cite{10577478,10379826}}. Unfortunately, current VR testing tools cannot properly characterize and represent the diverse interaction behaviors. For example, the state-of-the-art (SOTA) VR testing tool, \textit{VRGuide}, can only test virtual objects with the ``\textit{click}'' interaction.
Second, VR applications contain \textit{complex 3D virtual environments} with interactions with virtual objects, thereby introducing a vast exploration state space. Take \textit{EscapeGameVR}, a popular open-source VR gaming application, as an example, which contains 44 scenes, 8,256 GameObjects, and 1,377 C\# script files. 
Third, VR testing often requires completing a sequence of tasks, e.g., finding a key, then using the key to open a door, next turning a handle, and finally pressing a button to escape. It is non-trivial to accomplish these complex tasks since they involve diverse interactions and trigger either events or actions in a specific order. 
In summary, the diverse and intricate nature of the VR interactions, coupled with large-scale virtual spaces and complex task-completion sequences, makes it difficult to comprehensively and efficiently test VR applications in a systematic and repeatable manner.
\textbf{Our Approach.}  
To address the above challenges, we present \textit{VRExplorer} to thoroughly test Unity-based VR applications by conducting in-depth scene exploration and comprehensive interactions with virtual objects. Notably, we consider Unity-based VR applications mainly due to Unity's dominant role in VR/MR markets~\cite{venturebeat}. 
To tackle the first challenge mentioned above, we design the \underline{E}ntity, \underline{A}ction, and \underline{T}ask (\textsc{EAT}) framework for modeling VR interaction behaviors in a generic way. 
%We implement the EAT framework analogous to object-oriented programming (OOP). 
This hierarchical framework enables a reusable modeling process across VR applications developed by diverse Unity versions and interaction plugins, such as \textsc{XRIT}~\cite{XRIT_docu, Virtual_reality_toolkit_for_the_Unity_game_engine}, \textsc{SteamVR}~\cite{SteamVR,Using_Traditional_Keyboards_in_VR_SteamVR_Developer_Kit_and_Pilot_Game_User_Study,Steam_VR_Plugin}, and \textsc{MRTK}~\cite{MRTK,Introduction_to_the_Mixed_Reality_Toolkit}, thereby enhancing cross-project \textit{generalizability}. %and \textit{versatility}. 
To address the second challenge, we present the \textit{VRExplorer} agent built upon the \textsc{EAT} framework. Incorporating meticulously designed path-finding algorithms into Unity's NavMesh, the \textit{VRExplorer} agent can achieve autonomous navigation in 3D virtual environments. Moreover, the \textit{VRExplorer} agent can also systematically execute diverse interaction decisions on top of the Probabilistic Finite State Machine (PFSM). 
To address the third challenge, the proposed model-based approach transforms intricate VR task execution sequences into structured task models to enable systematic testing and automated execution.
%The testing process is elaborated in Section~\ref{subsection: VRExplorer}.

% to the best of our knowledge, \textit{the first} generic and versatile three-layer abstraction framework for modeling interaction tasks in VR applications, and 
% Specifically, for the first challenge, we abstract common interaction patterns from our dataset of diverse VR projects, which we elaborate in Section~\ref{subsection: Model Abstraction}.
% Based on this abstraction, we design \textsc{EAT} framework, which decomposes VR behavior modeling into three layers: the \underline{E}ntity Interface layer, \underline{A}ction Abstract Class layer, and \underline{T}ask Model layer. This layered design enables reusable modeling across applications built with different Unity versions and interaction plugins, such as \textsc{XRIT}, \textsc{SteamVR}, and \textsc{MRTK}, thereby enhancing cross-project generalizability and versatility, which we elaborate in Section~\ref{subsection: EAT Framework}.
% To address the second challenge, we present \textit{VRExplorer}, which is built upon the \textsc{EAT} framework. \textit{VRExplorer} incorporates Unity’s NavMesh system to support autonomous navigation in 3D environments and utilizes a Probabilistic Finite State Machine (PFSM) to model and execute diverse interaction paths in a systematic and repeatable way. The testing process is elaborated in Section~\ref{subsection: VRExplorer}.

\textbf{Evaluation.} 
% Experimental evaluation, as discussed in detail in Section~\ref{sec: Results}, on eleven diverse Unity-based VR projects shows that \textit{VRExplorer} consistently outperforms the SOTA approach \textit{VRGuide}, achieving significantly higher coverage and better efficiency. Specifically, \textit{VRExplorer} yields up to \textbf{96.5\%} improvement in EC and \textbf{61.5\%} in MC across a broad range of projects in RQ2. Furthermore, ablation results highlight the essential contributions of each interaction module, confirming the rationality of our design in RQ3. These findings demonstrate the generalizability and versatility of \textit{VRExplorer} for automated VR testing in real-world development scenarios.
%Before the evaluation of \textit{VRExplorer}, we statistically analyze the collected dataset projects (RQ1). The dataset analysis confirms its representativeness, as discussed in Section~\ref{sec:RQ1: Dataset Statistical Analysis}. %, with projects spanning 2 to 2,004 scripts, 256 to 237,725 lines of code (LOC), 21 to 5,303 scenes, 19 to 220,257 GameObjects, and 1 to 128 scenes, covering diverse Unity versions (2019.x to 2023.x). Our experimental evaluation demonstrates that \textit{VRExplorer} significantly advances automated testing for VR applications. 
To comprehensively evaluate the proposed \textit{VRExplorer}, we conduct extensive experiments on \hn{11} representative VR projects. The experimental results demonstrate that the proposed \textit{VRExplorer} outperforms the SOTA approach \textit{VRGuide} with average performance gains of \hn{122.8\% and 52.8\%} in terms of executable lines of code (ELOC) coverage and method (function) coverage, respectively. 
%and 43.9\% faster convergence in Group1 projects, while showing 96.5\% higher EC and 61.5\% higher MC in Group2 projects. 
Moreover, the ablation study on the five ablated variants of \textit{VRExplorer} also validates the necessity of all modules of the EAT framework. More importantly, \textit{VRExplorer} has successfully detected two previously unknown real-world bugs (i.e., one \textit{functional} bug and one \textit{non-functional} bug), which can nevertheless be detected by the baseline. 
Further, \textit{VRExplorer} has also successfully detected one previously-confirmed bug.
%These results collectively establish \textit{VRExplorer} as a generic approach to addressing the unique challenges of VR testing, demonstrating significant improvements in efficiency, generalizability compared to the SOTA approach, and ability to detect complicated real-world bugs.

\textbf{Main Contributions} are summarized as follows:
\begin{itemize}[leftmargin=12pt,itemindent=*] % 移除左侧缩进

   % \item We construct a dataset of 104 high-quality Unity-based open-source VR projects from GitHub and perform statistical analysis to ensure that it covers diverse testing scenarios and interaction patterns. This dataset serves as the foundation for abstraction modeling and testing evaluation.
    \item We design the \textsc{EAT} framework for modeling complex interaction behaviors and tasks in VR applications. To the best of our knowledge,  EAT is \textit{the first} generic three-layer abstraction framework for VR testing based on the object-oriented programming (OOP) paradigm. 
    
    \item We present \textit{VRExplorer}\footnote{%\textit{VRExplorer} is available at
\url{https://github.com/TsingPig/VRExplorer_Release}}, a novel model-based testing tool to achieve effective interactions with diverse virtual objects and the exploration of complex VR scenes.
   
    \item We evaluate \textit{VRExplorer} extensively on \hn{11} representative VR projects and demonstrate its consistent performance superior to the SOTA approach in terms of ELOC coverage, method coverage, and interactable object coverage while preserving high efficiency. Moreover, our \textit{VRExplorer} can also successfully detect real-world bugs in VR projects. %Ablation study further confirms the effectiveness of each component in our framework.
\end{itemize}

 % \vspace{-0.3cm}

\section{Background}
%\begin{itemize}[leftmargin=0pt,itemindent=*] % 移除左侧缩进
%\item 

\textbf{Interactions in Unity-based VR Applications.} 
A Unity-based VR application typically consists of multiple \emph{Scenes}, each of which is structurally represented by a \emph{hierarchy} of \emph{GameObjects}. These GameObjects represent all visible and interactive elements in the 3D virtual environment and can be composed of various components, such as meshes, scripts, and colliders. 
%Modern VR applications are predominantly developed using 
Unity provides VR application developers with a rich set of tools for implementing immersive interactions. As Unity's official framework, XRIT~\cite{XRIT_docu} can support common VR input modalities such as ray-based selection, direct grabbing, teleportation, and gesture recognition. These interactions are typically implemented using component-based scripts attached to objects, while often relying on trigger colliders or physics events. %, or animation states.
However, the implementation logic varies significantly across projects, especially when third-party packages are used, thereby increasing the complexity of generalizing automated testing across different VR projects.
%This diversity in interaction logic and object tagging increases the complexity of generalizing automated testing across different VR projects.

\textbf{Mono Scripts in Unity-based Application Development.} 
In Unity-based VR application development, \emph{Mono} scripts constitute the fundamental building blocks for implementing interactive behaviors. These C\# classes inherit properties from Unity's core \texttt{\small MonoBehaviour}~\cite{Mono_doc} base class, enabling them to leverage Unity's component-based architecture. Through the Unity Inspector, developers attach these scripts to GameObjects to create diverse objects. 
For example, a gun can be attached with a \texttt{\small XRGun} class inherited from \texttt{\small MonoBehaviour}, with properties referencing the bullet \textit{prefab}\footnote{Prefab is a reusable template~encapsulating GameObjects and components.}, \texttt{\small Shoot()}, etc.
% that shoots bullets 
%In Unity,, enabling efficient instantiation and modular design.
 %\item 

 % \textbf{Model-Based Testing in Software Engineering.}
 % Model-based testing (MBT)~\cite{Guided_stochastic_model-based_GUI_testing_of_Android_apps, Model_based_GUI_Testing_For_HarmonyOS_Apps, Search_Based_Automated_Play_Testing_of_Computer_Games:_A_Model-Based_Approach, Fastbot2} is a well-established approach in software engineering that uses abstract models to represent the desired behavior of a testing system. These models are typically FSM, extended finite state machines (EFSM), state transition diagrams, or other formal representations. MBT generates test cases based on the system's model, providing a systematic approach to exploring different scenarios that may not be easily covered through manual testing. By using these models, test engineers can automatically derive action sequences that are executed to validate system behavior under various conditions.

%\item 
\textbf{NavMesh-Based Navigation in Unity.}
Unity's NavMesh system~\cite{unity2020_navmesh, UnityNavMeshAgent, UnityNavMeshObstacle} provides a mechanism to traverse 3D environments using navigation meshes generated from static scene geometry. It supports obstacle avoidance, pathfinding, and dynamic updates, making it suitable for simulating players' movement. In VR testing, NavMesh can be leveraged to automate scene exploration by guiding a virtual agent through reachable areas. However, it only supports locomotion and requires external control logic to handle task-specific actions, such as interacting with objects or triggering events.

%\end{itemize}

\begin{figure*}[htbp] 
    \centering
    \includegraphics[width=0.95\textwidth]{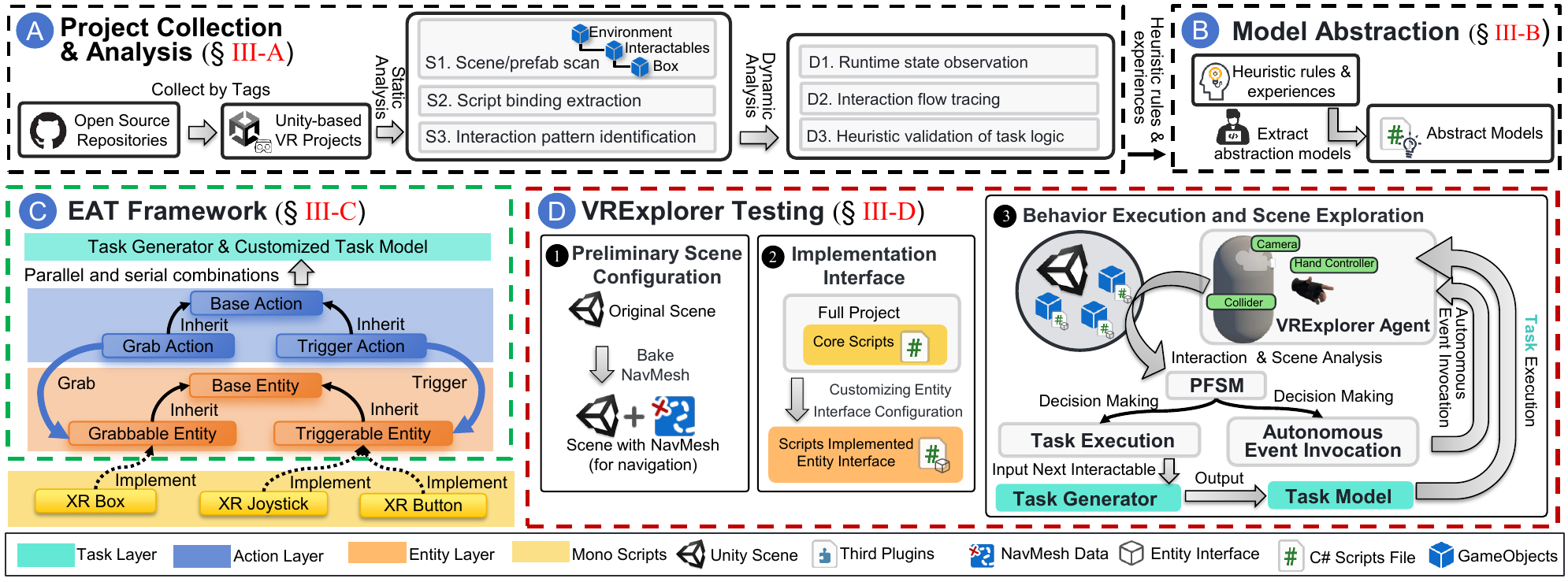}  
    \vspace{-0.26cm}
    \caption{\hn{Overview of \textit{VRExplorer}}}
    \label{fig:Overview of VRExplorer}
    \vspace{-0.4cm}
\end{figure*}

\section{Approach}
\label{sec:Approach}

\newcommand{\circlednumber}[1]{%
  \tikz[baseline=(char.base)]{%
    \node[circle, fill=black, inner sep=1.15pt, text=white] (char) {\footnotesize  #1};%
  }%
}
\definecolor{myblue}{HTML}{4472C4}
\newcommand{\circledalphabet}[1]{%
  \tikz[baseline=(char.base)]{%
    \node[circle,
          fill=myblue,
          inner sep=1.8pt,
          minimum width=1.2em, % 强制统一大小
          text=white,
          font=\sffamily\bfseries\small] (char) {\MakeUppercase{#1}};%
  }%
}

%\subsection{Overview}
%This section provides an overview of our approach, as shown in 
As shown in Fig.~\ref{fig:Overview of VRExplorer}, the proposed \textit{VRExplorer} works by first collecting and analyzing open-source VR projects in \S~\ref{subsection: Project Collection and Preliminary Analysis}. Then, it implements the EAT framework in \S~\ref{subsection: EAT Framework} after model extraction in \S~\ref{subsection: Model Abstraction}. 
%\S~\ref{subsection: EAT Framework} next elaborates on . 
Next, it conducts testing by \textit{VRExplorer} in \S~\ref{subsection: VRExplorer} based on the EAT framework, NavMesh, and PFSM.
%presents \textit{VRExplorer}, a model-based approach for automated VR testing based on the EAT framework, NavMesh, and PFSM.
%, designed to address generalizability and versatility issues in VR development.
%In Section~\ref{subsection: VRExplorer}, we introduce \textit{VRExplorer}, a model-based approach for automated VR application testing, which is based on the EAT framework, NavMesh and PFSM.

\subsection{Project Collection and Analysis}
\label{subsection: Project Collection and Preliminary Analysis}

\ZZY{Upd: 这段几乎算是重写了，为了回应ReviewB 的Comments 和Q1 针对 Project Analysis 和 Model Abstraction 部分说的不清楚的问题。主要说了一下我们采用了 "two-pass semi-automated analysis" 的方法，静态方面和动态方面分别详细说明。}

To comprehensively investigate interaction behaviors in VR, we collect open-source Unity VR projects with diverse applications, interactions, and scene types (details in \S~\ref{subsec:perf}).

\hn{
Based on this dataset, we perform a \emph{two-pass semi-automated analysis} --- static pass and dynamic pass---and then consolidate the results to construct Model Abstraction, including Object and Action abstractions (\S~\ref{subsection: Model Abstraction}). These abstractions are subsequently fed into the EAT framework (\S~\ref{subsection: EAT Framework}) and the PFSM-based testing workflow (\S~\ref{subsection: VRExplorer}).}

\hn{
\noindent\textbf{Static pass (S1--S3).}
\emph{S1: Scene/prefab scan.}
We parse scene hierarchies and prefabs to enumerate GameObjects with attached components, and prune \textit{non-interactable} infrastructure (e.g., static walls) %, lighting, and game managers
that are not connected to core interactions.
\emph{S2: Script binding extraction.}
We inspect \texttt{\small MonoBehaviour}-based scripts and their serialized fields to recover object--script bindings (e.g., references to targets). %prefabs, and anchor transforms).
\emph{S3: Interaction pattern identification.}
We analyze interaction-related APIs and callback signatures %(e.g., input- or physics-driven handlers, UnityEvent listeners, and library-specific interaction components) 
to fingerprint typical behaviors. %(e.g., grabbing, triggering, transforming). 
The interaction distribution will be detailed in \S~\ref{subsec:Implementation}.
% \emph{S4. Library-aware hints.}
% When third-party VR interaction libraries are present, we include their interaction components and event hooks in the static feature set.%
% \footnote{Our analysis does not depend on a specific library; components are treated as typed features discovered from project assemblies and serialized metadata.}
}

\hn{
\noindent\textbf{Dynamic pass (D1--D3).}
Static inspection cannot observe runtime object creation, dynamic component attachment, or goal-oriented task logic. 
To complement this effect, we conduct a lightweight but human-guided runtime analysis guided by gameplay flows and task objectives:
%where researchers play through and observe the execution frame by frame
\emph{D1: Runtime state observation.}
At each frame $t$, we record snapshots of active GameObjects $\mathcal{G}_t$ and their component sets $\mathcal{C}_t$. 
Comparing consecutive states ($\Delta\mathcal{G}_t$, $\Delta\mathcal{C}_t$), we identify dynamically instantiated objects %(e.g., via \texttt{\small Instantiate()}) 
and runtime-added components. %(e.g., via \texttt{\small AddComponent()}). 
%Manual observation ensures that short-lived or conditional objects, which may appear only in specific gameplay contexts, are not overlooked.
\emph{D2: Interaction flow tracing.}
Following execution evidence such as console outputs, function call stacks, and Unity event dispatches during gameplay,
%Rather than relying solely on static enumeration of callbacks, 
we trace when and how scripts are actually triggered (e.g., input events, collision handlers, and task-specific event chains). 
% This enables us to capture the true runtime entry points of interactions and their corresponding targets.
\emph{D3: Heuristic validation of task logic.}
Four experienced engineers heuristically reconstruct event sequences observed during gameplay to approximate these flows and validate whether each interaction contributes essentially to task progression or it is only incidental.
}

%hrough direct exploration and manual validation, we confirm whether an identified interaction is essential to task progression or merely incidental.  %(e.g., “grab key → open door → trigger scene transition”)

% \noindent 
% This human-in-the-loop runtime analysis is deliberately lightweight and complementary: automated logging provides raw evidence, while manual playthrough and reasoning ensure that dynamic, context-dependent interactions and goals are faithfully captured.

\subsection{Model Abstraction}
\label{subsection: Model Abstraction}

%After project analysis, we conduct model abstraction by generalizing VR interactable objects into 

\ZZY{Model Abstraction 部分新增了一大段落，回应ReviewB的针对建模步骤不清晰的Comments}

\hn{
\textbf{Heuristic Model Abstraction.} 
We summarize the heuristic experiences and rules extracted from the project analysis in order to test the framework's effectiveness and get a preliminary model abstraction. Specifically, during the project analysis process, we merge static and dynamic findings into abstraction models heuristically and classify common VR interaction behaviors into abstract actions based on OOP principles, as shown in Part B in Fig.~\ref{fig:Overview of VRExplorer}.
When testing a real-world project, engineers can absolutely use the model we had constructed, while also experiencing core gameplay and performing the two-pass manual analysis as described in \S~\ref{subsection: Project Collection and Preliminary Analysis} if they have customized testing demands.
New interaction patterns are either matched with existing types or newly defined via new Action interfaces automatically. While execution is fully automated, abstraction still requires human-in-the-loop analysis due to the complexity of VR interactions and the limited domain-specific knowledge of existing tools. 
Objects are annotated with interactable features (e.g., \textit{Grabbable}, \textit{Triggerable}, and \textit{Transformable}); interactions are lifted into abstract actions. %(Table~\ref{tab:Example of the Abstact Action}).
This consolidated model (i) seeds the \emph{Entity Interface Layer} in EAT by proposing interface candidates (e.g., \texttt{\small IGrabbable} and \texttt{\small ITriggerable}), (ii) provides action-level semantics to the \emph{Action Class Layer}, and (iii) surfaces callable functions and UnityEvents to populate the task %and \emph{exploration} nodes in the PFSM behavior space 
(\S~\ref{subsection: VRExplorer}).
% By construction, dynamically instantiated objects and runtime-attached components are not missed: any $\Delta\mathcal{G}_{t}$ or $\Delta\mathcal{C}_{t}$ is integrated before abstraction.
}

\begin{table}[htbp]
\caption{Example of Generalizing VR Interactable Objects}
%with Interactable Features
\vspace*{-0.2cm}
\centering
\resizebox{0.725\linewidth}{!}{
\begin{tabular}{lc}
\toprule
\textbf{VR Interactable Objects} & \textbf{Interactable Features} \\ %\hline
{\color[HTML]{333333} Box, Coin, etc.} & \textit{\color[HTML]{333333} Grabbable} \\ %\hline
Button & \textit{Triggerable} \\ %\hline
Joystick & \textit{Transformable} \\ %\hline
Gun, Cigarette Lighter, etc. & \textit{\color[HTML]{333333} Grabbable} and \textit{Triggerable} \\
 \bottomrule
\end{tabular}
}
\label{tab:Example of the Abstact Feature}
\vspace*{-0.2cm}
\end{table}
%\item
\textbf{Abstraction of Interactable Objects.} We first generalize VR interactable objects into abstract objects containing interactable features. Table~\ref{tab:Example of the Abstact Feature} gives an example of VR interactable objects with interactable features, which can determine interaction properties.  
%These features categorize the interactable objects and highlight their core attributes. 
For instance, objects like boxes and coins can be typically classified as \textit{Grabbable}, meaning that they can be picked up, while objects like buttons and joysticks are considered \textit{Triggerable} with two attributes \textit{triggering} and \textit{triggered}. Notably, some objects may possess multiple interactable features. For instance, a gun or a cigarette lighter possesses both \textit{Grabbable} and \textit{Triggerable} attributes, representing complex interactions to support not only grabbing but also acting on its function, such as shooting or turning on a button.

%\end{itemize}

%\vspace*{-0.3cm}
\begin{table}[htbp]
\caption{Generalizing Interactions into Abstract Actions}
\vspace*{-0.2cm}
\centering
\resizebox{0.725\linewidth}{!}{
\begin{tabular}{lc}
\toprule
\textbf{VR Interaction Behaviors} & \textbf{Abstract Actions} \\%\hline
{\color[HTML]{333333} Move Around, Jump, Fly, etc.} & \textit{\color[HTML]{333333} Move Action} \\%\hline
Press, Pull, Open/Close, Turn On/Off, etc. & \textit{Trigger Action} \\%\hline
Grab, Throw & \textit{Grab Action} \\%\hline
Move/Scale/Rotate Objects & \textit{Transform Action} \\
\bottomrule
\end{tabular}
}
\label{tab:Example of the Abstact Action}
\vspace*{-0.2cm}
\end{table}
%\begin{itemize}[leftmargin=0pt,itemindent=*] % 移除左侧缩进
%\item 

\textbf{Abstraction of Actions.} We then classify VR interaction behaviors into abstract actions. 
Table~\ref{tab:Example of the Abstact Action} shows an example of a VR interactive scene, in which pressing a button, closing or opening a door, turning on or off a lamp, and pulling a joystick,  can be conceptualized as compositions of a continuous process with a following event. Specifically, the player's action, pulling a joystick, can be classified into a continuous pulling process and an event triggered after pulling is over. We classify a behavior with such characteristics into \textit{Trigger Action}.

\subsection{EAT Framework}
\label{subsection: EAT Framework}
% 在模型抽象的基础上，为了更好的解决通用性和泛化性问题，我们提出了EAT框架，这是一个的三层框架，包括实体接口层（Entity interface layer）、行为抽象类层（Action abstract class layer）和任务层（Task Model layer）。
Based on the model abstraction in \S~\ref{subsection: Model Abstraction}, we propose a three-layer framework called EAT, as shown in Part C in Fig.~\ref{fig:Overview of VRExplorer}. 
%which stands for \underline{E}ntity Interface layer, \underline{A}ction Abstract Class layer, and \underline{T}ask Model layer, for better generalizability and versatility in VR application testing.

% 该层由继承自MonoBehaviour类的核心VR交互逻辑类组成，通常在Inspector窗口中挂载到VR可交互对象上。例如，一把能够发射子弹的枪可以挂载XRGun类，这个类继承自MonoBehaviour，并包含诸如如对子弹预制体的引用指针变量和Shoot()函数等。
%\begin{itemize}[leftmargin=0pt,itemindent=*] % 移除左侧缩进
%\item 
\textbf{Entity Interface Layer.} Based on the interactable objects' abstraction, we define specific interfaces to encapsulate their corresponding attributes.
%of VR interactable objects. 
Particularly, we define a base interface, called \texttt{\small BaseEntity}, which includes the \texttt{\small Transform} property to represent the object's position, rotation, and scale, as well as the \texttt{\small Name} property. Thereafter, all other entity interfaces inherit this base interface. For example, an interface \texttt{\small Triggerable} includes two attributes \textit{triggering} and \textit{triggered} with the \texttt{\small Enum} property. The Entity Interface layer provides customized interfaces tailored to specific features required by the Action layer. Notably, complex interactable features can be efficiently represented through multi-inheritance by multiple interfaces, thereby allowing for higher flexibility and modularity in defining VR interactions. 
% For example, a gun class \texttt\{\small XRGun}, which possesses both \textit{Grabbable} and \textit{Triggerable} features, can be implemented by inheriting from \texttt{\small MonoBehaviour} while simultaneously implementing the \texttt{\small IGrabbable} and \texttt{\small ITriggerable} interfaces. This allows the gun object to be both grabbed and triggered during VR interactions without additional script components.

% \lstdefinestyle{mystyle}{
%     language=[Sharp]C,            % 指定 C# 语言
%     basicstyle=\ttfamily\footnotesize, % 代码字体
%     keywordstyle=\color{blue},    % 关键词颜色
%     commentstyle=\color{gray},    % 注释颜色
%     stringstyle=\color{red},      % 字符串颜色
%     numbers=left,                 % 显示行号
%     numbersep=5pt,                % 行号间距
%     numberstyle=\tiny\color{gray}, % 行号格式
%     breaklines=true,              % 自动换行
%     frame=single,                 % 代码块外框
%     backgroundcolor=\color{black!5}, % 代码背景色
%     captionpos=b,                 % 标题位置
%     tabsize=4,                    % Tab 长度
% }
% \begin{figure}[htbp] 
%     \centering
%     \includegraphics[width=\linewidth]{Figure/Fig_Example_of_Entity_Layer.pdf}  
%     \caption{Overview of \textit{VRExplorer}}
%     \label{fig:Example of }
% \end{figure}

% % 使用自定义代码样式
% \begin{lstlisting}[style=mystyle]
% class XRGun: MonoBehaviour, IGrabbable, ITriggerable {
%     // Implementation of grabbable and triggerable features
% }
% \end{lstlisting}

\begin{figure}[t]
    \centering
    % 图1在下方
    % 图2和图3并排显示
    \vspace{-0.3cm}
    \begin{minipage}[b]{0.30\linewidth}  % 设置图2的宽度
        \centering
        \includegraphics[width=\textwidth]{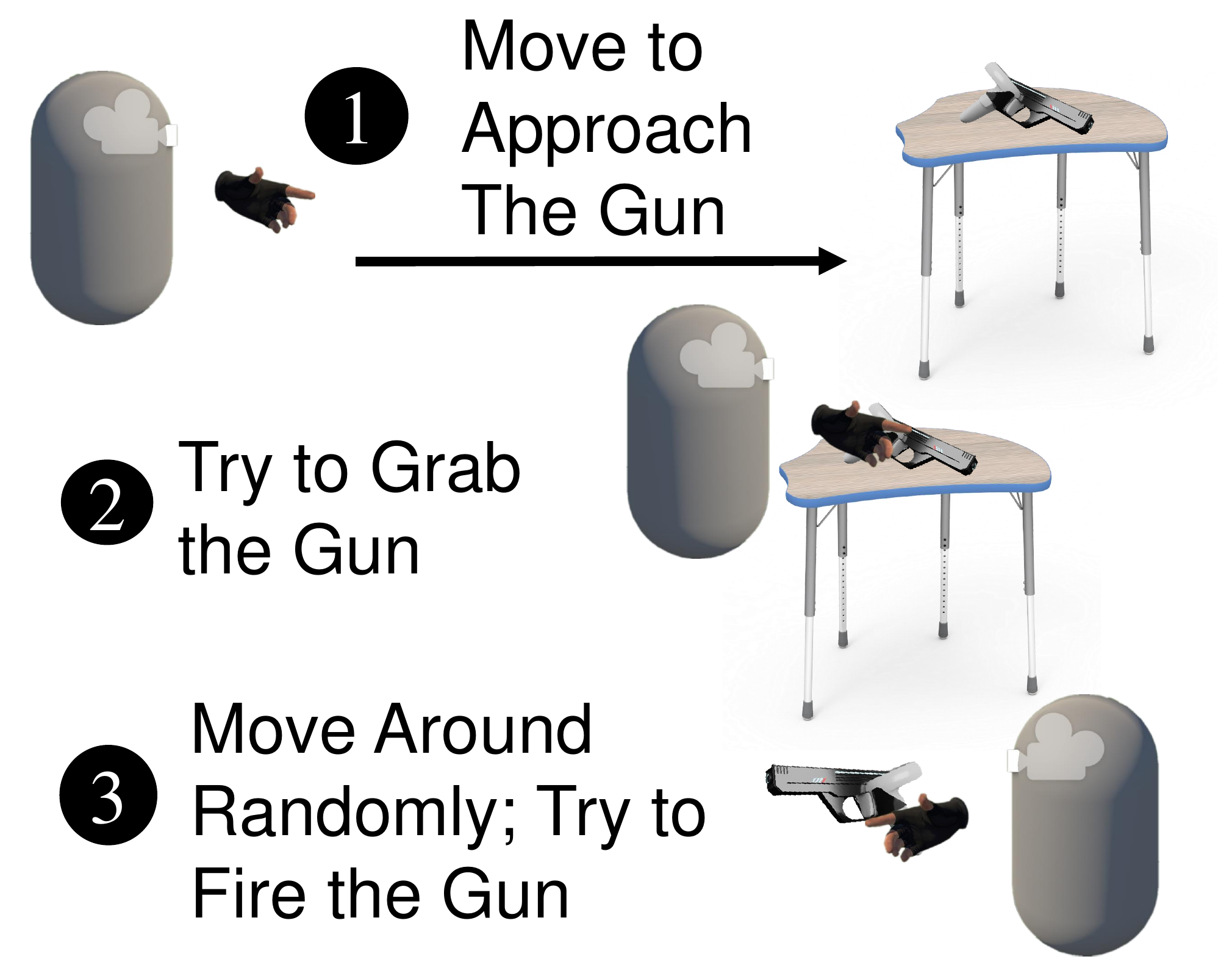}
        \subcaption{GASG Process}
        \label{fig:a Task Instance Model}
    \end{minipage}
    \hfill  % 添加水平间隔，使两图之间有空隙
    \begin{minipage}[b]{0.65\linewidth}  % 设置图3的宽度
        \centering
        \includegraphics[width=\textwidth]{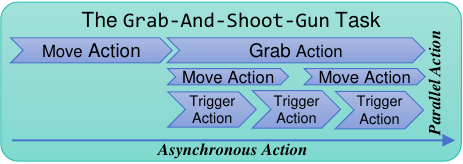}
        \subcaption{Action Model of GASG task}
        \label{fig:a Task Instance Action}
    \end{minipage}
       \vspace{0.1cm}  % 图1和图2、图3之间的间隔
    \begin{minipage}[b]{\linewidth}
        \centering
        \includegraphics[width=0.95\textwidth]{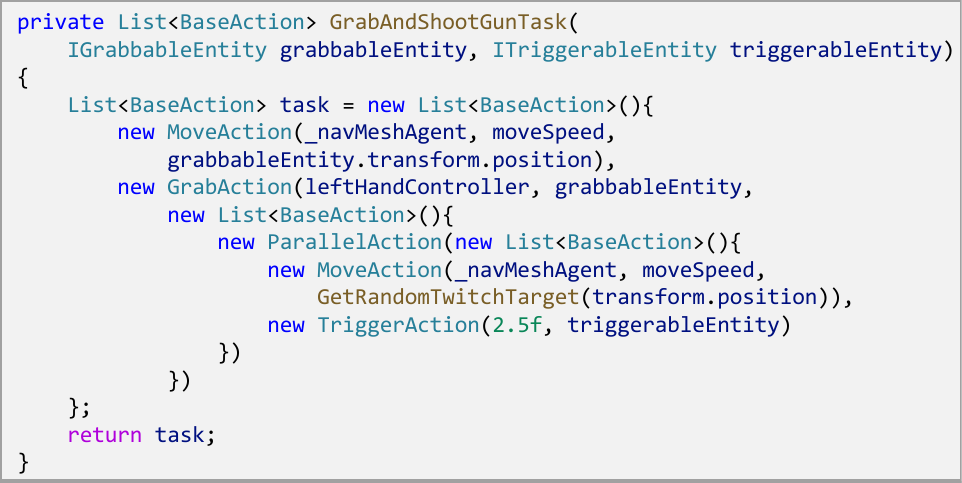}
        \subcaption{C\# Code Snippet of GASG task}
        \label{fig:a Code Snippet for a Task Instance}
    \end{minipage}
    \vspace*{-0.55cm}
        \caption{Task Instance of \texttt{\small Grab-And-Shoot-Gun }}
    \label{fig:group_of_task_instance}
    \vspace*{-0.4cm}
\end{figure}

% 该层基于前文提到的行为抽象，将具有类似特征的动作归入同一动作类别。我们定义了一个抽象基类BaseAction，其中包含一个虚拟异步方法（Asynchronous Method）Execute()。所有其他行为类均继承自该基类。行为层能够与实体层交互，并为VRExplorer提供功能性API接口，模拟真实玩家的交互动作。
%\item
\textbf{Action Class Layer.} We then extract behaviors with identical characteristics into the same action. 
We first define an abstract base class, namely \texttt{\small BaseAction} with a \textit{virtual} asynchronous \texttt{\small Execute()} method. All other action classes inherit this base class. The Action layer can interact with the Entity Interface Layer and provide functional APIs for \textit{VRExplorer} to simulate real players' interactive actions. For example, \textit{Trigger Action} can be concreted into the \texttt{\small TriggerAction} class, which consists of an asynchronous method \texttt{\small Triggering()} and a synchronous method \texttt{\small Triggered()}.

% 任务是多个行为的串行和并行组合。并行的多个行为之间在同一时刻同时执行各自的异步方法，串行的多个行为之间有先后顺序，必须等待前一个行为执行完成才能继续后续行为。任务层包含任务生成器和多个预定义的任务，任务生成器通过接受 MonoBehaviour 实例或者实体数组 BaseEntity[] 生成对应的任务。预定义的任务是我们根据前期对项目的分析，抽象了一些常用的任务模型，例如 xx 所示，我们定义了 Grab-And-Shoot-Gun 任务，如图b所示，这个任务分为三个步骤，靠近桌子，拿起枪，边走边射击。其对应的模型如图c所示，横轴代表着异步任务的时间轴，首先是一个 Move Action，接着是一个 Parallel Action，其内部的三个行为之间彼此并行。对应的代码如图c所示。

%\item
\textbf{Task Model Layer.} 
A composition task consists of multiple \textit{parallel} and \textit{sequential} actions. Parallel action executes asynchronous methods simultaneously, while sequential actions follow a strict order to complete the previous action before proceeding to the next.  
The Task Model layer includes a task generator and multiple predefined VR interaction tasks. The task generator creates tasks by accepting either a \texttt{\small MonoBehaviour} instance or an array of entities (\texttt{\small BaseEntity[]}). We derive the predefined tasks from commonly used task models.  
%via our initial project analysis. 
Fig.~\ref{fig:group_of_task_instance} depicts an example of the \texttt{\small Grab-And-Shoot-Gun} (GASG) task, which consists of three steps: approaching the table, picking up the gun, and shooting while walking in Fig.~\ref{fig:group_of_task_instance}(\subref{fig:a Task Instance Model}). Fig.~\ref{fig:group_of_task_instance}(\subref{fig:a Task Instance Action}) elaborates on the corresponding action model, where the horizontal axis represents the timeline of asynchronous actions starting with \textit{Move Action}, followed by \textit{Parallel Action}, within which three actions execute concurrently. The corresponding code is also shown in Fig.~\ref{fig:group_of_task_instance}(\subref{fig:a Code Snippet for a Task Instance}).

%\end{itemize}

\subsection{\textit{VRExplorer} Testing}
\label{subsection: VRExplorer}

As shown in Part D in Fig.~\ref{fig:Overview of VRExplorer}, \textit{VRExplorer}'s testing process works in three sequential steps:
% \circlednumber{1} Preliminary Scene Configuration,
% \circlednumber{2} Implementation Interface, and
% \circlednumber{3} Scene Exploration with Behavior Execution.

% \begin{algorithm}[t]
% \caption{Entity Implementation Example: LaunchProjectile }
% \label{alg:launch-projectile}
% \begin{algorithmic}[1]
% \STATE \textbf{class} LaunchProjectile \textbf{implements} ITriggerableEntity, IGrabbableEntity
% \STATE \COMMENT{\textcolor{blue}{\#region Entity}}
% \STATE \qquad \textbf{property} TriggeringTime $\gets$ 1.5
% \STATE \qquad \textbf{property} Name $\gets$ ``Gun''
% \STATE \qquad \textbf{property} Grabbable:
% \STATE \qquad \qquad \textbf{if} component exists \textbf{then return} it
% \STATE \qquad \qquad \textbf{else} add new Grabbable component
% \STATE \qquad \textbf{method} Triggerring() \textbf{do} nothing
% \STATE \qquad \textbf{method} Triggerred() \textbf{calls} Fire()
% \STATE \qquad \textbf{method} OnGrabbed() \textbf{do} nothing
% \STATE \COMMENT{\textcolor{blue}{\#endregion}}
% \STATE \textbf{field} projectilePrefab: GameObject
% \STATE \textbf{field} startPoint: Transform
% \STATE \textbf{field} launchSpeed: float
% \STATE \textbf{method} Fire():
% \STATE \qquad Instantiate projectile and apply forces
% \STATE \textbf{method} ApplyForce(rigidbody):
% \STATE \qquad Calculate and apply physics forces
% \end{algorithmic}
% \end{algorithm}

\begin{figure}[t]
    \centering
    \begin{minipage}[t][3cm][t]{0.49\linewidth}
        \centering
        \includegraphics[height=2.5cm]{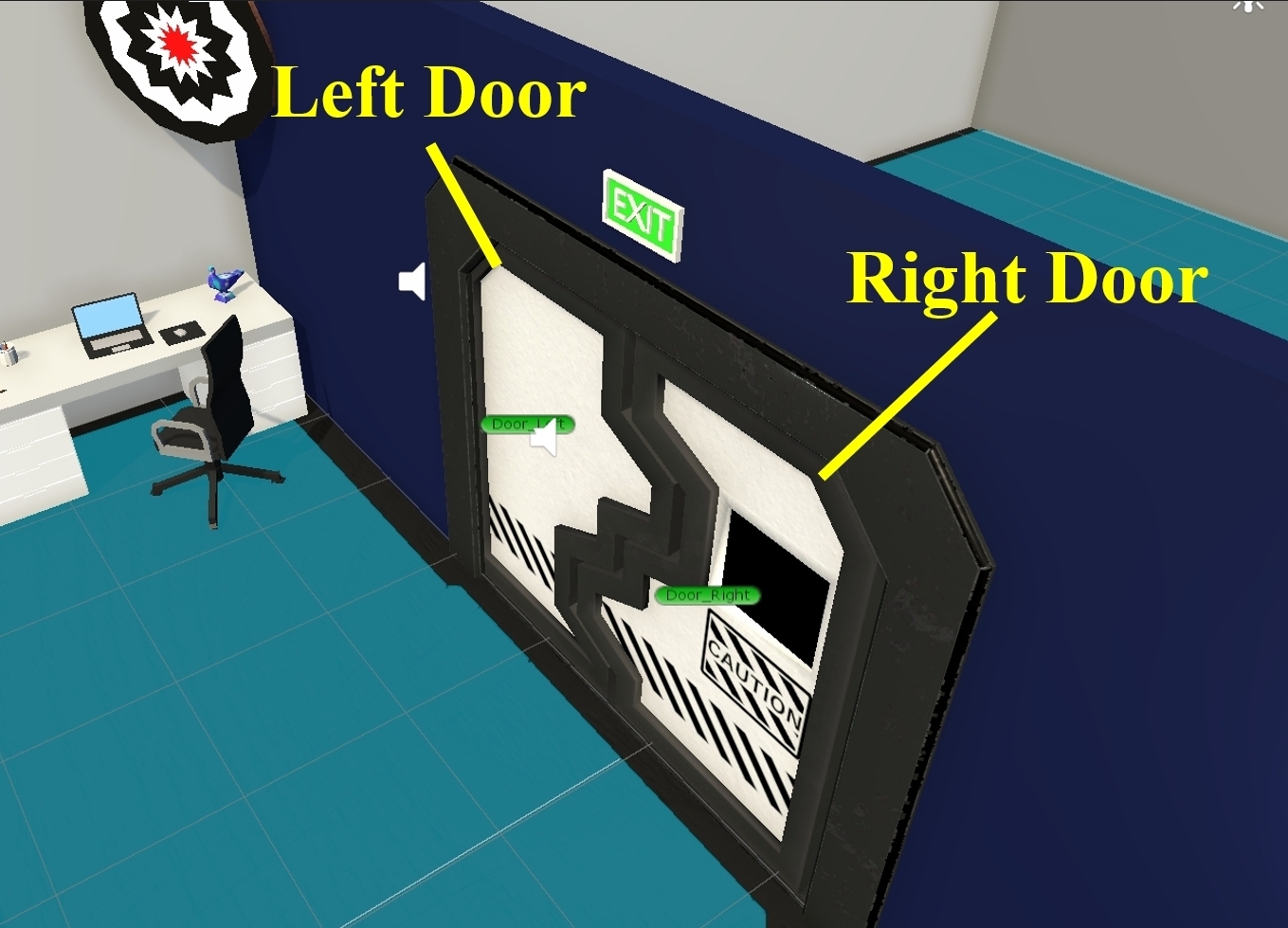}
        \subcaption{\scriptsize{Doors Closed (NavMesh separated)}} 
        \label{fig:door_closed}
    \end{minipage}
    \hfill
    \begin{minipage}[t][3cm][t]{0.49\linewidth}
        \centering
        \includegraphics[height=2.5cm]{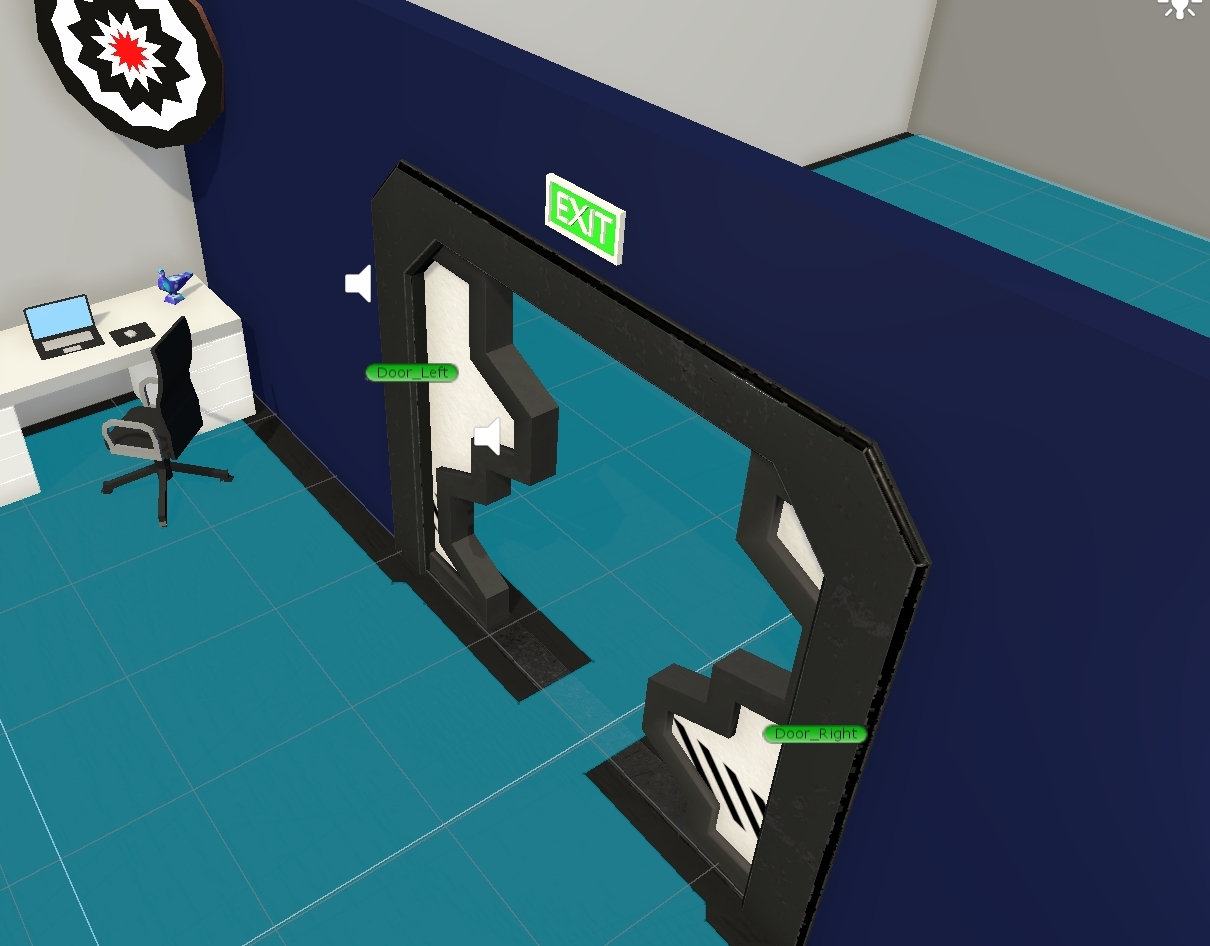}
        \subcaption{\scriptsize{Doors Opened (NavMesh connected)}}
        \label{fig:door_opened}
    \end{minipage}
    \vspace*{-0.1cm}
    \caption{Dynamic doors with NavMesh Obstacle \small{(Carve enabled)}}
    \label{fig:Doors}
    \vspace*{-0.3cm}
\end{figure}

%\subsubsection{Preliminary Scene Configuration} 
\circlednumber{1} \textit{Preliminary Scene Configuration}, before scene exploration, we facilitate the agent's navigation capability based on Unity's NavMeshAgent~\cite{UnityNavMeshAgent} via NavMesh baking. We first configure terrain objects, such as floors and walls, to be static. Regarding those objects that may dynamically move, such as doors, we attach the NavMesh Obstacle~\cite{UnityNavMeshObstacle} component with the enabled Carve option, thereby allowing them to dynamically modify the NavMesh. Fig.~\ref{fig:Doors} depicts an example of opening and closing \textit{dynamic} doors. 
%do not need to be set as static. followed by NavMesh baking. Objects that may dynamically move, such as doors, do not need to be set as static. However, they should have 

%\subsubsection{Preliminary Scene Configuration}
% 由于导航部分使用Unity提供的 NavMeshAgent 系统，在场景配置时，需要将地板、墙壁等地形物体设置为静态，并进行导航网格烘焙。场景中可能会开启的门等物体不需要设置静态，但需要挂载上 \texttt{NavMesh Obstacle} 组件，并开启 Carve 选项，使其能够动态的切割导航网，正如图x所示

\circlednumber{2} \textit{\hn{Implementation Interface}}, 
\label{subsubsec:entity}
%\textbf{Entity Interface Layer Implementation.} 
on top of the EAT framework, implementing \textit{interactable} interfaces is the core to tackle two challenges: (1) lack of generalizability caused by diverse Unity versions and the fragmentation of the VR development ecosystem; and (2) intricate scene exploration and VR interactions. Using interfaces to encapsulate implementation details can enable testing for a diversity of VR applications.
% \begin{figure}[htbp] 
%     \centering
%     \includegraphics[width=0.9\linewidth]{Figure/Fig_Code_Snippet_of_Launch_Projectile_after_the_Entity_layer_implementation.pdf}  
%     \caption{Code Snippet of Launch Projectile after the Entity  layer implementation}
%     \label{fig:Code Snippet of Launch Projectile after the Entity  layer implementation}
% \end{figure}
% 实现接口是 \textit{VRExplorer} 基于EAT框架，来解决不同开发框架、不同类型VR应用的通用性测试的核心。通过接口屏蔽具体实现细节，使其能够测试多种多样的VR应用

% 为了能够覆盖并检测交互函数的可靠性，测试者需要对待测试程序集中与核心VR交互逻辑相关的 Mono层脚本选择和定制合适的 Entity层接口，并实现具体的接口函数。如图x所示，图中橙色的部分表示的是实现接口时额外增加的代码。

% \begin{procedure}[h]

% \caption{Entity Interface Layer Implementation Example.}%\\ \textit{\textbf{\small Note:}{\footnotesize The implementation of entity interface code lines is marked blue.}}}
% \label{proc:entity}
% \footnotesize
% \begin{algorithmic}[1]
% \STATE \textbf{class} \texttt{\footnotesize XRGun} \textbf{implements} \texttt{\footnotesize ITriggerable}, \texttt{\footnotesize IGrabbable}:

% \STATE \qquad \textcolor{blue}{\textbf{property} TriggeringTime $\gets$ 1.5}
% \STATE \qquad \textcolor{blue}{\textbf{property} Name $\gets$ ``Gun''}
% \STATE \qquad \textcolor{blue}{\textbf{property} Grabbable:}
% \STATE \qquad \qquad \textcolor{blue}{\textbf{if} component exists \textbf{then return} it}
% \STATE \qquad \qquad \textcolor{blue}{\textbf{else} add new Grabbable component}
% \STATE \qquad \textcolor{blue}{\textbf{method} \texttt{\scriptsize Triggerring()} \textbf{do} nothing}
% \STATE \qquad \textcolor{blue}{\textbf{method} \texttt{\scriptsize Triggerred()} \textbf{calls} \texttt{\scriptsize Fire()}}
% \STATE \qquad \textcolor{blue}{\textbf{method} \texttt{\scriptsize OnGrabbed()} \textbf{do} nothing}

% \STATE \qquad \textbf{field} projectilePrefab: GameObject
% \STATE \qquad \textbf{field} startPoint: Transform
% \STATE \qquad \textbf{field} launchSpeed: float
% \STATE \qquad \textbf{method} \texttt{\scriptsize Fire()}:
% \STATE \qquad \qquad Instantiate projectile and apply forces
% \STATE \qquad \textbf{method} \texttt{\scriptsize ApplyForce(rigidbody)}:
% \STATE \qquad \qquad Calculate and apply physics forces
% \end{algorithmic}
% \end{procedure}

\begin{procedure}[h]
\caption{\small{Entity Interface Layer Implementation Example.}}
\label{proc:entity}
\scriptsize
\begin{algorithmic}[1]
\STATE \textbf{class} \texttt{\scriptsize XRGun} \textbf{implements} \texttt{\scriptsize ITriggerable}, \texttt{\scriptsize IGrabbable}:
\STATE \qquad \textcolor{blue}{\textbf{property} TriggeringTime $\gets$ 1.5}
\STATE \qquad \textcolor{blue}{\textbf{property} Name $\gets$ ``Gun''}
\STATE \qquad \textcolor{blue}{\textbf{property} Grabbable:}
\STATE \qquad \qquad \textcolor{blue}{\textbf{if} component exists \textbf{then return} it}
\STATE \qquad \qquad \textcolor{blue}{\textbf{else} add new Grabbable component}
\STATE \qquad \textcolor{blue}{\textbf{method} \texttt{\scriptsize Triggerring()} \textbf{do} nothing}
\STATE \qquad \textcolor{blue}{\textbf{method} \texttt{\scriptsize Triggerred()} \textbf{calls} \texttt{\scriptsize Fire()}}
\STATE \qquad \textcolor{blue}{\textbf{method} \texttt{\scriptsize OnGrabbed()} \textbf{do} nothing}
\STATE \qquad \textbf{field} projectilePrefab: GameObject
\STATE \qquad \textbf{field} startPoint: Transform
\STATE \qquad \textbf{field} launchSpeed: float
\STATE \qquad \textbf{method} \texttt{\scriptsize Fire()}:
\STATE \qquad \qquad Instantiate projectile and apply forces
\STATE \qquad \textbf{method} \texttt{\scriptsize ApplyForce(rigidbody)}:
\STATE \qquad \qquad Calculate and apply physics forces
\end{algorithmic}
\end{procedure}

To ensure coverage and verify the reliability of interaction methods, test engineers need to select and customize the appropriate Entity layer interfaces for the Mono scripts related to the core VR interaction logic in the target project with the corresponding interface functions implemented. 
% Procedure~\ref{proc:entity} shows an example of Entity Interface Layer Implementation.
% As shown in~\ref{fig:Code Snippet of Launch Projectile after the Entity  layer implementation}, the orange sections indicate the additional code introduced during interface implementation.
Take Procedure~\ref{proc:entity} as an example\footnote{The implementation of the entity interface's code lines is marked blue.}, in which a gun class \texttt{\small XRGun} possessing both \textit{Grabbable} and \textit{Triggerable} features, can be implemented by inheriting \texttt{\small MonoBehaviour} while simultaneously implementing the \texttt{\small IGrabbable} and \texttt{\small ITriggerable} interfaces. This design allows the gun object to be both \textit{grabbed} and \textit{triggered} during VR interactions without additional script components. 
This approach allows \texttt{\small XRGun} to seamlessly integrate multiple interaction capabilities (i.e., \textit{grabbable} and \textit{triggerable}). Leveraging interface composition, we can flexibly define and extend object behaviors without modifying the base class hierarchy, thereby promoting code reusability and achieving modular design in the VR interaction system.

\begin{figure}[b] 
%  \vspace{-0.3cm}
    \centering
    \includegraphics[width=0.9\linewidth]{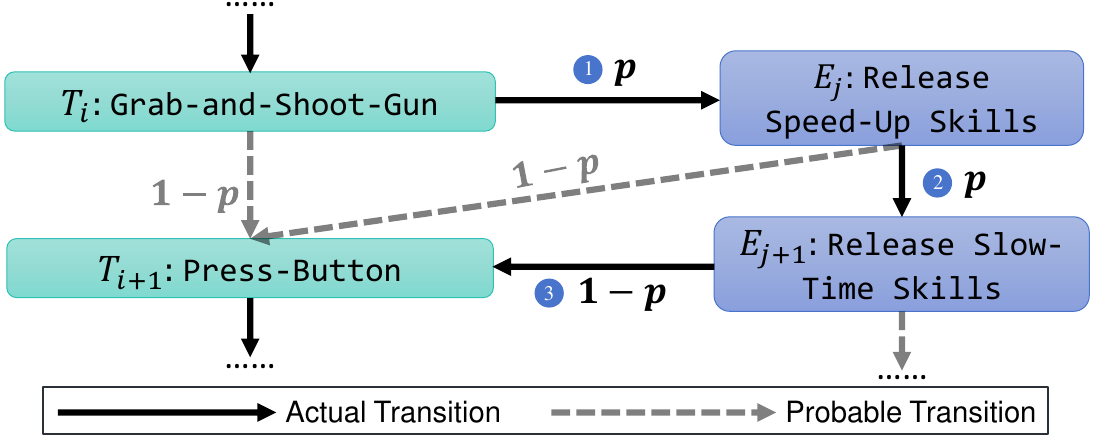}  
       \vspace{-0.1cm}
    \caption{\hn{Example of PFSM State Transition in a VR Scene.}}
    \vspace{-0.4cm}
    \label{fig:An instance of State Transition}
\end{figure}

\circlednumber{3} \textit{Scene Exploration with Behavior Execution}.
\label{subsubsec:Scene Exploration with Behavior Execution}
after completing the scene configuration and implementing interactable interfaces, \textit{VRExplorer} automatically performs scene exploration and VR interactions by executing the corresponding behaviors. We design two types of behaviors: \textit{task execution} and \textit{{autonomous event invocation}}. When executing tasks, \textit{VRExplorer} perceives the scene and inputs \texttt{\small MonoBehaviour} instances or an array of entities (\texttt{\small BaseEntity[]}) into the task generator to obtain and execute the corresponding tasks. Notably, players in VR environments not only interact with objects but also invoke autonomous events, such as casting skills to gain acceleration buffs.  
%autonomous events reveal that 
%\begin{itemize} [leftmargin=0pt,itemindent=*]

%\item 
\textbf{PFSM.}
To comprehensively cover such testing cases, \textit{VRExplorer} maintains a configurable list of \texttt{\small UnityEvents}, allowing test engineers to easily customize the functions to be covered by configuring them in the Inspector before testing. These two types of behaviors constitute the behavior space, where each behavior can be regarded as a node. The behavior space itself forms a directed graph composed of these nodes. For behavior decision-making, \textit{VRExplorer} employs PFSM to determine transitions between nodes, thereby defining the topology of the directed graph. We use $\mathcal{T} = \{T_1, T_2, \dots, T_N\}$ to denote the set of all task states, where $T_i$ represents the $i$-th task state in the execution sequence, with $i \in \{1, 2, \dots, N\}$. Similarly, we use $\mathcal{E} = \{E_1, E_2, \dots, E_M\}$ to denote the set of exploration states, where $E_j$ represents the $j$-th exploration state in the execution sequence, with $j \in \{1, 2, \dots, M\}$. At each decision point in PFSM, a variable determines the transition direction. The probability of transitioning to an exploration state ($E_j$) is denoted by $p$ while the probability of transitioning to a task state ($T_i$) is $(1 - p)$. Fig.~\ref{fig:An instance of State Transition} depicts an example of state transition in a VR scene, where \texttt{\small Speed-Up Skills} node and \texttt{\small Slow-Time Skills} node are two example nodes of an autonomous event adding buffs to the player, while GASG node and Press-Button (PB) node are two example nodes of the task. Those four nodes are independent of each other, while the PFSM is responsible for decision-making and state transitions among them. Table~\ref{tab:State Transition Table of PFSM} lists all the state transitions.

\begin{table}[t]
\caption{State Transition Table of PFSM}
\vspace{-0.25cm}
\centering
\footnotesize
  \renewcommand{\arraystretch}{0.8}
\begin{tabular}{llll}
\toprule
\textbf{Current State} & \textbf{Next State} & \textbf{Condition} & \textbf{Probability} \\
\midrule
$T_i$ & $T_{i+1}$ & Default & $1-p$ \\
$T_i$ & $E_j$ & Default & $p$ \\
$E_j$ & $E_{j+1}$ & Default & $p$ \\
$E_j$ & $T_{i+1}$ & Default & $1-p$ \\
$T_i$ & $T_{i+1}$ & $\mathcal{E}$ exhausted & 1.0 \\
$E_j$ & $E_{j+1}$ & $\mathcal{T}$ exhausted & 1.0 \\
\bottomrule
\end{tabular}
\label{tab:State Transition Table of PFSM}
\vspace{-0.4cm}
\end{table}

%\item 
\textbf{Path-finding for Navigation.}
We implement two algorithms: (i) a Greedy algorithm, which follows a local optimization strategy based on the shortest path principle, and (ii) a Backtracking algorithm with Pruning, which searches for globally optimal solutions. Since the Greedy algorithm significantly reduces time complexity, fulfilling the real-time requirements of VR testing, we mainly adopt it in our experiments (\S~\ref{sec: Discussion} presenting a comparison of the two algorithms).

In summary, \textit{VRExplorer} continuously obtains scene information, decides the current behavior to execute, and receives feedback after performing the behavior. This process is repeated until all interactable objects and events are covered.

%\end{itemize}

\section{Evaluation of \textit{VRExplorer}}
\label{sec:evaluation}

%We then evaluate the proposed VRExplorer by comparing the SOTA baselines on eleven representative projects. 

\begin{table}[b]
    \centering
    \caption{\hn{Statistical Metrics of Collected 102 Projects.}}
\vspace{-0.3cm}
    \resizebox{\linewidth}{!}{
    \begin{tabular}{lrrrrrrr}
    \toprule
     Metric & Mean & Variance & Min & Q1 & Median & Q3 & Max \\
    \midrule
    Scripts & 166.45 & 110,953.52 & 2 & 26.25 & 71.50 & 130.25 & 2,004 \\
    LOC     & 19,910.85 & 1,851,165,810.13 & 256 & 2,707.25 & 6,129.00 & 13,762.75 & 237,725 \\
    Files   & 902.07 & 902,718.22 & 21 & 301.25 & 620.50 & 1,090.00 & 5,303 \\
    Scenes  & 18.12 & 455.06 & 1 & 4.00 & 8.00 & 30.25 & 128 \\
    \bottomrule
    \end{tabular}
    }
    \label{tab:Statistical Metrics of Dataset Projects}
\end{table}

%\subsection{Implementation and Environment}
\subsection{Implementation}
\label{subsec:Implementation}
\begin{figure}[t] 
    \centering
    \includegraphics[width=0.9\linewidth]{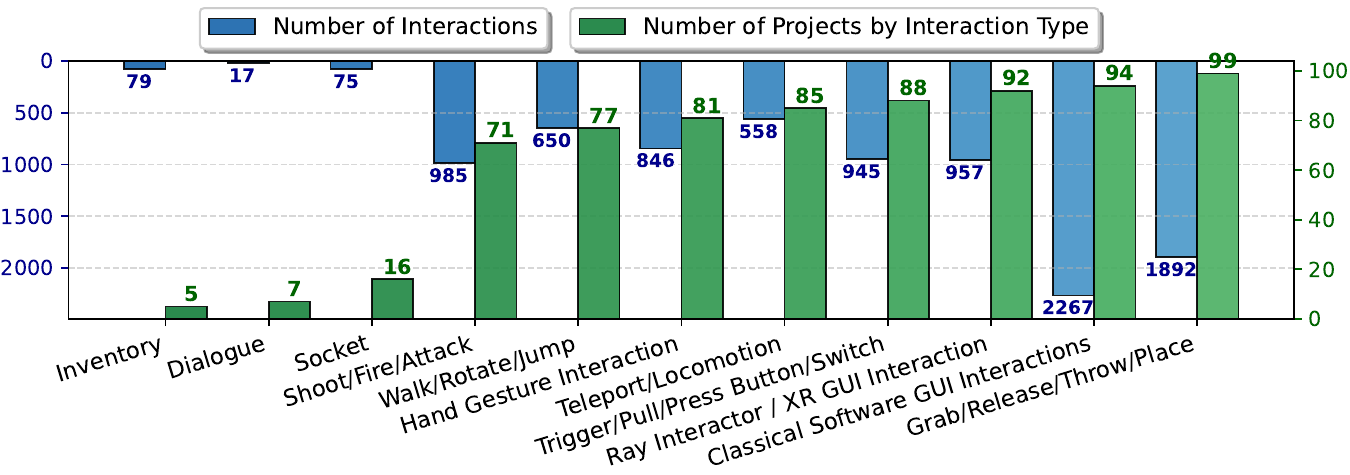}  
    \caption{\hn{Distribution of Interactions in Dataset.}}
    \label{fig:Number of Interactions and Number of Projects by Interaction Type}
    \vspace{-0.35cm}
\end{figure}

\hn{
\textbf{Project Collection and Analysis.} We implemented a crawler to collect Unity-based VR GitHub projects with keywords like “VR”, “AR”, “unity”, and “xr”. From 971 initially filtered projects, we manually performed quality checks (e.g., compilation errors and version conflicts), consequently retaining 102 high-quality projects.
We then analyze the structural and interaction characteristics of all projects in the dataset to assess their representativeness and complexity. 
Table~\ref{tab:Statistical Metrics of Dataset Projects} shows varied project sizes. While most projects are relatively small (medians containing 71.5 scripts, 6,129 LOC, 620 files, and 8 scenes), a few of them with large sizes markedly raise the medians. Fig.~\ref{fig:Boxplots of Scripts, LOC, Files, and Scenes Metrics in Dataset Projects} plots their distribution. 
Fig.~\ref{fig:Number of Interactions and Number of Projects by Interaction Type} indicates that the dataset includes both prevalent interactions (e.g., grabbing, GUI operations, and teleportation) and infrequent ones (e.g., inventory and dialogue). Together, these interactions cover all major interaction tasks identified in \cite{A_Taxonomy_of_Interaction_Techniques_for_Immersive_Augmented_Reality_based_on_an_Iterative_Literature_Review}, thereby highlighting the constructed dataset’s representativeness and diversity.
}

\begin{figure}[t] 
    \centering
    \vspace{-0.15cm}
    \includegraphics[width=0.9\linewidth]{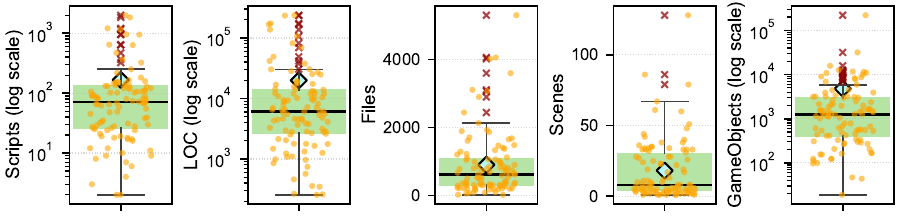}  
    \caption{\hn{Boxplots of scripts, LOC, files, and scenes.}} %across projects
    \label{fig:Boxplots of Scripts, LOC, Files, and Scenes Metrics in Dataset Projects}
\vspace{-0.35cm}
\end{figure}

\hn{
We then perform the two-pass analysis (\S~\ref{subsection: Project Collection and Preliminary Analysis}), summarizing heuristic rules for subsequent modeling.}
%and experiences as drafted documentation

\hn{
\textbf{Implementation of Model Abstraction.}  
Based on the above analysis, we extract abstract models using the derived heuristics. The implementation comprises 6 scripts in the Action layer, 4 scripts in the Entity layer, and 4 pre-defined Mono scripts. Additionally, 5 pre-defined task models orchestrate these components, resulting in more than 1,600 lines of code in total. Manual exploration per project ranged from 30 to 90 minutes, depending on the engineer’s proficiency.
}

\textbf{Implementation of the EAT Framework.} 
% 具体而言，我们根据Core Code，在Entity layer中，我们从base interface \texttt{\small BaseEntity}派生出 \texttt{\small Grabbable}, \texttt{\small Transformable} and \texttt{\small Triggerable}接口，然后再Action layer中从abstract base class \texttt{\small BaseAction}派生出\texttt{\small GrabAction}, \texttt{\small MoveAction}, \texttt{\small ParallelAction}, \texttt{\small TransformAction}, and \texttt{\small TriggerAction}子类，并分别按照其具体行为，重写\textit{virtual} method \texttt{\small Execute()}，并将常见的任务模型代码在\textit{VRExplorer}中实现。
We derive \texttt{\small Grabbable}, \texttt{\small Transformable}, and \texttt{\small Triggerable} interfaces from the base interface \texttt{\small BaseEntity} in the Entity layer. In the Action layer, we define \texttt{\small GrabAction}, \texttt{\small MoveAction}, \texttt{\small ParallelAction}, \texttt{\small TransformAction}, and \texttt{\small TriggerAction} as the subclasses of the abstract base class \texttt{\small BaseAction}. Each subclass overrides the \texttt{\small Execute()} method to implement its corresponding behavior. Thereafter, we provide the implementations of common task models as the code-level compositions of actions, e.g., the PB and GASG tasks.
%, as shown previously in Fig.~\ref{fig:group_of_task_instance}(\subref{fig:a Code Snippet for a Task Instance}). 

%\textbf{Implementation of the Entity Interface layer.} 
For each project, we analyze the interactable objects in scenes and all the C\# scripts referenced by the objects, and select the codes corresponding to the core VR interaction functionality, referred to as \hn{\textit{Core Code}. 
Tool libraries, third-party code, and other non-interaction scripts are excluded. This selection serves two purposes: (i) evaluation of testing performance via Assembly-Definition files, and (ii) flexible customization and configuration for different VR projects.  
Core Code is a \textit{heuristic criterion} useful for any type of VR applications (even Unity-based or non-Unity-based).}

\begin{figure}[t] 
  \vspace{-0.3cm}
    \centering
    \includegraphics[width=0.9\linewidth]{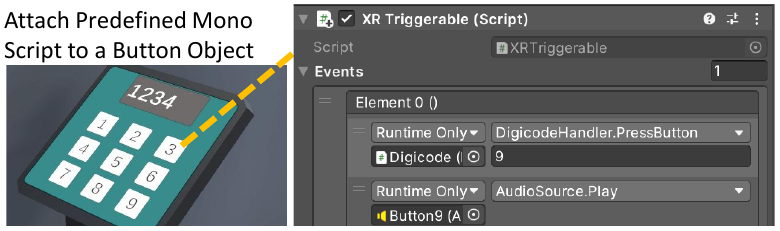}  
    \caption{Predefined Mono Script Component \texttt{\small XRTriggerable}}
    \label{fig:Pre-defined Mono Script Component XRTriggerable}
    \vspace{-0.3cm}
\end{figure}

We then modify classes in \textit{Core Code} to implement the interfaces of the Entity Interface layer. To minimize additional workload for developers, we provide a set of predefined scripts inherited from \texttt{\small MonoBehaviour} for projects developed using the \textsc{XRIT} framework. These commonly used Mono scripts already implement the corresponding interfaces. Developers can simply attach the predefined scripts to target objects and configure the methods under test via \texttt{\small UnityEvent}~\cite{UnityEvent2025} in the Unity Inspector (see Fig.~\ref{fig:Pre-defined Mono Script Component XRTriggerable}).
\hn{This design allows easy extension to diverse interactions across VR applications while introducing minimal extra effort for developers, and ensures that the Core Code is fully testable and easily configurable.}

\textbf{Implementation of \textit{VRExplorer}.}
% 我们为\textit{VRExplorer} agent的GameObject添加了NavMeh Agent用于在导航网格上寻路，并为他添加了控制器（类似玩家的手）用于和物体交互。为了感知环境，我们实现了\texttt{\small EntityManager}这个类，并且由他注册场景信息、并为\textit{VREplorer}的场景分析提供APIs接口。我们通过条件分支语句的状态判断实现PFSM用于决策，场景分析得到的数据传递给PFSM进行决策。我们为\textit{VRExplorer} agent 的 task executation 提供了 \texttt{\small Task Generator}负责将对应的输入转换成需要的任务模型。
We equip the \textit{VRExplorer}'s GameObject with a NavMeshAgent component for navigation on NavMesh with the attached controller component (like a player's hands) to enable interactions with objects. To support scene perception, we implement the \texttt{\small EntityManager} class, which is responsible for registering scene information and providing APIs for scene analysis within \textit{VRExplorer}. Then, we implement PFSM to support conditional branching for decision-making, where PFSM's inputs are derived from scene analysis. Next, we implement a \texttt{\small TaskGenerator} component to support task execution by translating relevant inputs into the corresponding task model.

%\item 
\textbf{Environment.} We implement and evaluate \textit{VRExplorer} with the comparison of other baselines (in \S~\ref{subsec:baseline}) on a computer with AMD Ryzen 7 5800H with Radeon Graphics CPU, 32GB RAM, and NVIDIA GeForce RTX 3060 GPU (6GB) Graphics card. This computer is installed with Windows 11 and the same Unity version as that used in all evaluated projects.

%\end{itemize}

%\begin{itemize}[leftmargin=0pt,itemindent=*] % 移除左侧缩进
%\item 
\textbf{System Configuration Parameters.}
Since the proposed \textit{VRExplorer} simulates a player to explore the VR scene, the \textit{move speed} (MS) and the \textit{turn speed} (TS) (also called \textit{rotation speed} in~\cite{VRGuide}) greatly affect the testing performance. 
%In particular, we let the center of the scene be the initial position. 
%; this setting is the most appropriate. 
Regarding the move speed, excessive speed can introduce physical and interactive issues (e.g., test tools may unintentionally move out of a platform due to inertia), although a faster speed may generally improve performance. Therefore, to balance efficiency with practicality, we choose MS = 6 m/s and TS = 60 deg/s as the standard parameters for \textit{VRGuide} and \textit{VRExplorer} in all subsequent experiments. \S~\ref{sec: Discussion} discusses the rationale for these settings. 
Since these parameters are chosen to be the same for all test tools, they are not repeated in detail in the subsequent experimental results. An experiment is terminated when a test tool reaches convergence.
%\item 
%\textbf{Other parameters.}
In the PFSM model, the probability of transitioning to the next state is decided by parameter $p$, which is set to 0.5, indicating the same propensity for both state transitions.

%\end{itemize}

\begin{table}[t]
  \centering
  
  \caption{Quantitative Metrics of Selected VR Projects}
  \vspace{-0.20cm}
  
  \label{tab:vr_projects}
  \resizebox{\linewidth}{!}{%
  \renewcommand{\arraystretch}{0.9}
  \begin{threeparttable}
  \begin{tabular}{@{}ll r r r r r r@{}}
    \toprule
    & 
    \textbf{Projects} & 
    \textbf{\# of Scripts} & 
    \textbf{LOC} & 
    \textbf{\# of Files} & 
    \textbf{Scenes} & 
    \textbf{\# of GOs} & 
    \textbf{Version} \\
    \midrule
    %\multicolumn{7}{l}{\textbf{Group1}} \\
    %\cmidrule(l){2-7}
    \multirow{3}{*}{\rotatebox[origin=c]{90}{ Group 1}}
    & \texttt{unity-vr-maze} & 158 & 25,261 & 212 & 1 & 278 & 5.x \\
    & \texttt{UnityVR} & 150 & 24,858 & 330 & 3 & 124 & 2019.x \\
    & \texttt{UnityCityView} & 182 & 28,335 & 446 & 34 & 1,194 & 2019.x \\
    \cmidrule(l){1-8}
    \multirow{8}{*}{\rotatebox[origin=c]{90}{ Group 2}}
    %\multicolumn{7}{l}{\textbf{Group2}} \\
    & \texttt{Parkinson-VR}\tnote{1} & 275 & 38,437 & 968 & 33 & 1,566 & 2019.x \\
    & \texttt{VGuns} & 81 & 10,900 & 848 & 36 & 1,653 & 2020.x \\
    & \texttt{EE-Room}\tnote{2} & 88 & 4,450 & 1,063 & 8 & 1,517 & 2020.x \\
    & \texttt{EscapeGameVR} & 91 & 6,659 & 1,377 & 44 & 8,256 & 2021.x \\
    & \texttt{VRChess} & 160 & 26,591 & 414 & 4 & 280 & 2021.x \\
    & \texttt{VR-Basics} & 62 & 2,677 & 724 & 5 & 2,143 & 2021.x \\
    & \texttt{VR-Room} & 65 & 3,660 & 679 & 2 & 414 & 2022.x \\
    & \texttt{VR-Adventure} & 11 & 260 & 91 & 2 & 288 & 2022.x \\
    \bottomrule
  \end{tabular}%
   \begin{tablenotes}
  \item[1]Parkinson-VR stands for Parkinson-App-Virtual-Reality.
  \item[2]EE-Room stands for Edutainment-Escape-Room.
  \end{tablenotes}
  \vspace{-0.45cm}
\end{threeparttable}
  }%

 % \footnotesize \textit{\textbf{Notes:}} Group1's project comes from part of the dataset given in~\cite{VRGuide}, the project's version is older and they include click events, which can be tested by both \textit{VRGuide} and \textit{VRExplorer}, allowing for a direct comparison of efficiency and performance; Group2's project comes from our collection, the version is updated, however, \textit{VRGuide}  cannot be effectively tested, so only its coverage can be compared.
  
\end{table}

\subsection{Metrics} With reference to~\cite{Navigating_Mobile_Testing_Evaluation:A_Comprehensive_Statistical_Analysis_of_Android_GUI_Testing_Metrics, VRGuide, Towards_Agent_Based_Testing_of_3D_Games_using_Reinforcement_Learning, Aplib_Tactical_Agents_for_Testing_Computer_Games}, we consider ELOC coverage, method coverage, interactable objects coverage, and convergence time as evaluation metrics:
\begin{itemize}[leftmargin=0pt,itemindent=*] % 移除左侧缩进
\item \textbf{ELOC Coverage (\hn{EC}).}
Since ELOC mainly focuses on code lines containing executable programs, EC measures the percentage of these executable lines with the exclusion of blank lines, comments, and declaration statements during testing. We adopt \textit{Code Coverage}~\cite{CodeCoverage_Github%, AboutCodeCoverage
}, an official tool provided by Unity, to automatically record the EC of C\# scripts. Running in Unity Editor Mode, the testing tool exports the coverage data as a historical report in \texttt{\small .html} and a summary in \texttt{\small .xml}.

\item \textbf{Method Coverage (\hn{MC}).} Besides EC, we also consider MC to evaluate performance, quantifying the percentage of testing methods (functions) that have been invoked during testing. 
    
\item \textbf{Interactable Object Coverage (\hn{IOC}).} We adopt IOC to fairly compare \textit{VRExplorer} with SOTA baseline \textit{VRGuide}~\cite{VRGuide}. Slightly different from~\cite{VRGuide}, in which interactable objects only include objects that can receive mouse ``click'' events, we further extend interactable objects to all objects that support user interaction in the VR environment. % \textit{triggered}, \textit{grabbed}, \textit{moved}, and \textit{transformed}. 
%for example, a gun that can be grabbed and fired, a key that can be used to open a door, or a remote controller for operating a toy car. 
These interactable objects are identified by analyzing and confirming the \texttt{\small MonoBehaviour} scripts associated with the scene objects. 
   
\item \textbf{Convergence Time.} We also evaluate the efficiency of the proposed approach. Particularly, we consider \textit{convergence time} to measure how fast a tool can reach the \textit{converged state}, which is defined as the status at which the testing tool no longer seeks additional scene exploration. 
%, indicating a lack of further exploration intent or capability. 
The convergence time refers to the amount of time taken for the testing tool to reach the converged state from the initial state.
\end{itemize}

%while in real VR scenarios, we think it is more reasonable to define them as
\subsection{Baselines}
\label{subsec:baseline}
Given the unique characteristics of VR applications, such as complex VR interactions and the fragmented nature of the VR development ecosystem, current automated testing tools and Android application testing frameworks are not suitable for VR application testing. To the best of our knowledge, \textit{VRGuide}~\cite{VRGuide} is the SOTA testing approach for VR applications. Besides \textit{VRGuide}, \textit{VRTest}~\cite{VRTest} is a previous version, which nevertheless has inferior performance to \textit{VRGuide} in terms of both MC and IOC (only pointer click events received). 
%Different from them, our \textit{VRExplorer} 

\section{Experimental Results}
\label{sec:result}

\begin{table*}[h]
\caption{Results on Projects in Group 1}
\vspace{-0.2cm}
\centering
\resizebox{0.95\linewidth}{!}{
\renewcommand{\arraystretch}{0.8}
\begin{tabular}{ccccccc}  % ← 正确为6列
\toprule
\textbf{Projects} & \textbf{Approaches}  &   \textbf{EC (\%)} & \textbf{MC (\%)} & \textbf{IOC(\%)} & \textbf{Convergence Time Cost (s)} & \textbf{\# of Interactable Objects} \\
\midrule
\multirow{2}{*}{\texttt{unity-vr-maze}} 
 & \textit{VRGuide} & 66.53 & 70.59 & 94.29 & 145.0 & \multirow{2}{*}{35}\\
 & \textit{VRExplorer} & \colorbox{lightergray}{\parbox[c][1.0mm][c]{2cm}{\centering 81.67 (+22.8\%)}} & \colorbox{lightergray}{\parbox[c][1.0mm][c]{2cm}{\centering  82.35 (+16.7\%)}} & \colorbox{lightergray}{\parbox[c][1.0mm][c]{2cm}{\centering 100.00 (+6.1\%)}} & \colorbox{lightergray}{\parbox[c][1.0mm][c]{2cm}{\centering 81.4 (-43.9\%)}} \\ \midrule

\multirow{2}{*}{\texttt{UnityCityView}} 
 & \textit{VRGuide} & 67.66 & 78.38 & 60.00 & 45.0& \multirow{2}{*}{15}  \\
 & \textit{VRExplorer} & \colorbox{lightergray}{\parbox[c][1.0mm][c]{2cm}{\centering 92.22 (+36.3\%)}} & \colorbox{lightergray}{\parbox[c][1.0mm][c]{2cm}{\centering 100.00 (+27.6\%)}} & \colorbox{lightergray}{\parbox[c][1.0mm][c]{2cm}{\centering 100.00 (+66.7\%)}} & \colorbox{lightergray}{\parbox[c][1.0mm][c]{2cm}{\centering 89.3 (+98.4\%)}} \\ \midrule

\multirow{2}{*}{\texttt{UnityVR}} 
 & \textit{VRGuide} & 64.81 & 84.62 & 100.00 & 8.8 & \multirow{2}{*}{3}\\
 & \textit{VRExplorer} & \colorbox{lightergray}{\parbox[c][1.0mm][c]{2cm}{\centering 75.93 (+17.1\%)}} & \colorbox{lightergray}{\parbox[c][1.0mm][c]{2cm}{\centering 92.31 (+9.1\%)}} & \colorbox{lightergray}{\parbox[c][1.0mm][c]{2cm}{\centering 100.00}} & \colorbox{lightergray}{\parbox[c][1.0mm][c]{2cm}{\centering 7.7 (-12.5\%)}}  \\
\bottomrule
\end{tabular}
}
\label{tab:Results of RQ2 on Group1}
\vspace{-0.3cm}
\end{table*}

%\subsection{Research Questions}
We conduct extensive experiments to evaluate \textit{VRExplorer} and aim to investigate three research questions (RQs):
\begin{itemize}[leftmargin=0pt,itemindent=*] % 移除左侧缩进
% \item \textbf{RQ1:} How representative and comprehensive is the collected dataset in reflecting the diversity and complexity of VR application testing scenarios and interaction patterns?
\item \textbf{RQ1:}  How does \textit{VRExplorer} perform with comparison of the existing SOTA approach in diverse VR projects? 
%(RQ1.1)?  Additionally, how does \textit{VRExplorer} perform in terms of coverage performance, specifically regarding MC, EC, and interactable objects coverage in a VR scene (RQ1.2)? 
% Additionally, how does \textit{VRExplorer} perform in terms of generalizability and versatility?
\item \textbf{RQ2:} How do different modules contribute to the performance of \textit{VRExplorer}? 

\item \textbf{RQ3:} Can \textit{VRExplorer} detect real-world VR bugs?
\end{itemize}

\subsection{\textit{VRExplorer} Performance}\label{subsec:perf}

\begin{figure}[t]
    \centering

        \begin{subfigure}[t]{\linewidth}
        \centering
        \includegraphics[width=0.9\linewidth]{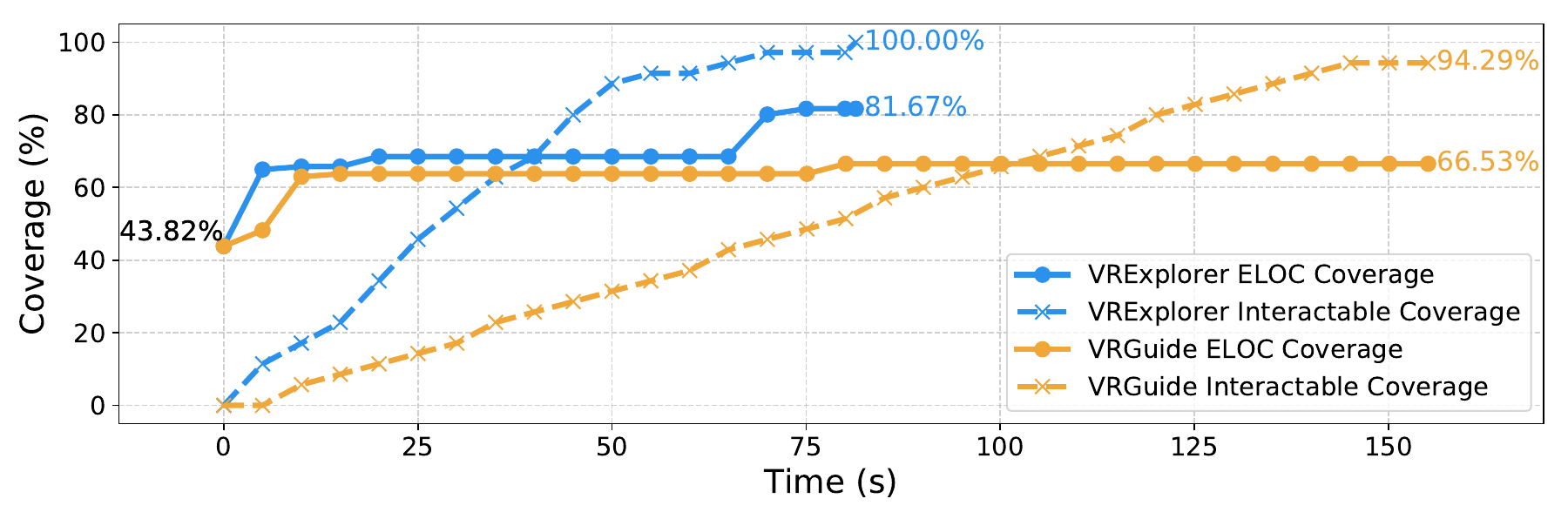}
        \vspace{-0.30cm}
        \caption{\scriptsize{Coverage versus Time in Project \texttt{unity-vr-maze}}}
        \label{fig:Coverage Performance of unity-vr-maze}
    \end{subfigure}

    \begin{subfigure}[t]{\linewidth}
        \centering
        \includegraphics[width=0.9\linewidth]{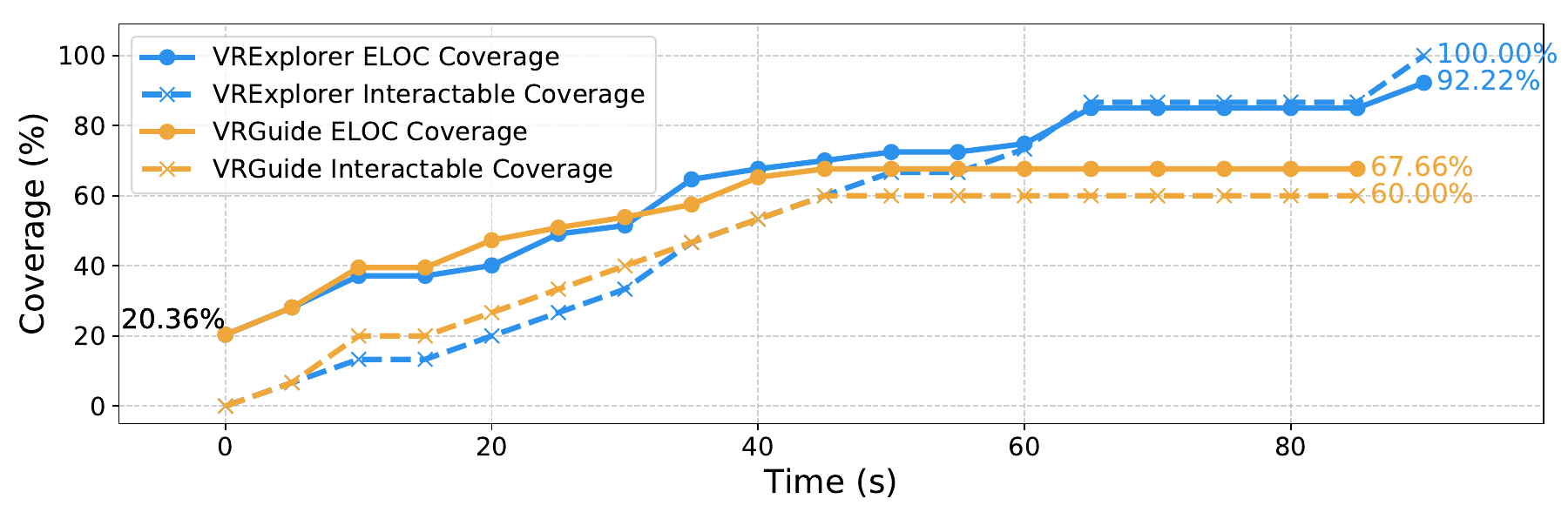}
               \vspace{-0.30cm}
        \caption{\scriptsize{Coverage versus Time in Project \texttt{UnityCityView}}}
        \label{fig:Coverage Performance of UnityCityView}
    \end{subfigure}

    \begin{subfigure}[t]{\linewidth}
        \centering
        \includegraphics[width=0.9\linewidth]{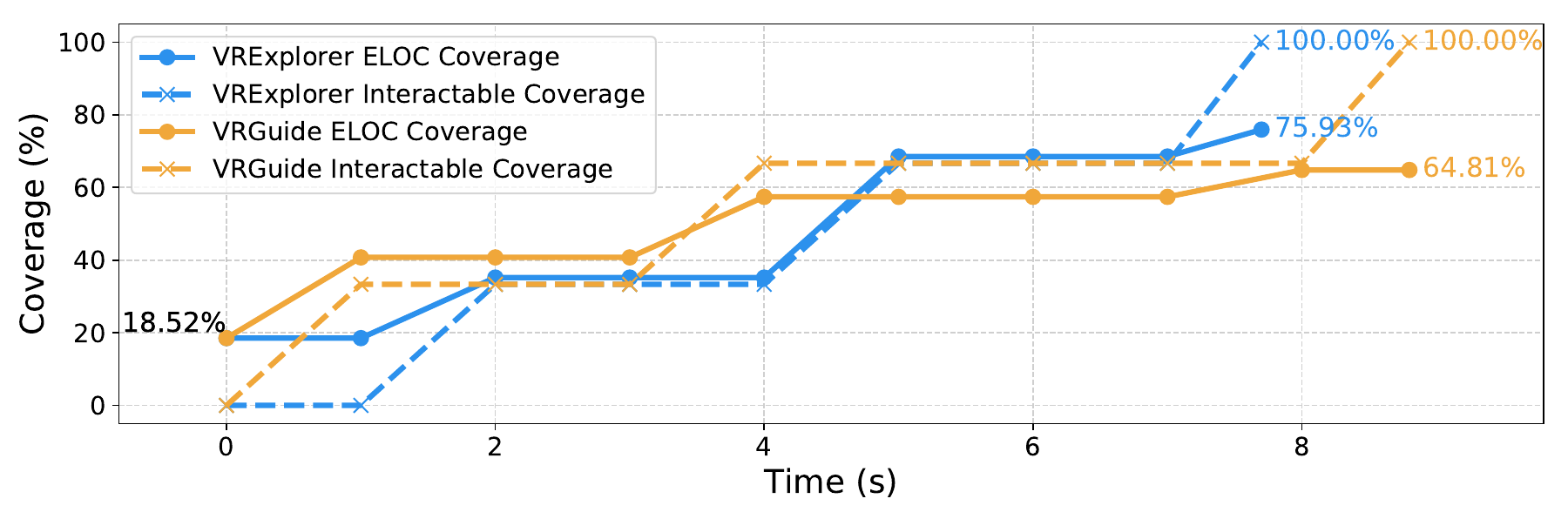}
               \vspace{-0.30cm}
        \caption{\scriptsize{Coverage versus Time in Project \texttt{UnityVR}}}
        \label{fig:Coverage Performance of UnityVR}
    \end{subfigure}
    
    %\vspace{0.5em} % 调整子图之间的间距
    
       \vspace{-0.2cm}
    \caption{EC versus time during the testing process~in~Group 1.}
     \vspace{-0.25cm}
        %\texttt{\small Awake()}) runs immediately, regardless of interactions.}}
    % The initial EC starts from the same value across all the projects because initialization code (e.g., \texttt{\small Start()} and \texttt{\small Awake()}) executes immediately, independent of interactions.}

    \label{fig:Results of RQ2 on Group1}

    % \smallskip

    %     \footnotesize \textbf{\textit{Note:}} The initial EC starts from the same value across all VR projects because initialization code (e.g., \texttt{\small Start()} and \texttt{\small Awake()}) executes immediately, independent of interaction.
    
\vspace{-0.55cm}
\end{figure}

%\begin{itemize}[leftmargin=0pt,itemindent=*] % 移除左侧缩进
%\item 

%\end{itemize}
% 修改：说明评估每个项目需要at least 5 hours，总量500小时比较大；并且项目之间有代表性，都评估比较冗余；因此我们寻求子集，选择有代表性的项目评估
% Evaluating each Unity project requires setting up a runnable environment locally, performing manual code analysis, and implementing customized interfaces. 
% On average, the preliminary configuration of the evaluation of a single project takes at least 5 hours, which means that it would take more than 500 hours to evaluate the entire data set.
% Given the substantial effort required and the fact that many projects in the dataset share similar interaction patterns and scene structures, conducting a full evaluation would be prohibitively time-consuming and redundant. Therefore, in order to balance workload and representativeness, we opted for a representative subset that covers diverse interaction patterns, application types, and Unity versions. 

\textbf{Constructing VR Projects for Evaluation.} 
To evaluate the proposed \textit{VRExplorer}, we construct a project dataset\footnote{List of constructed project dataset can be found at \url{https://github.com/TsingPig/VRExplorer/blob/main/Artifacts/Evaluated_Repo_Url.md}}, consisting of \hn{11} representative VR projects, as summarized in Table~\ref{tab:vr_projects}. This project dataset can be divided into two groups: (1) Group 1, in which we select the most complex three projects also being included in \textit{VRGuide} \cite{VRGuide}, and (2) Group 2, into which we introduce the other \hn{eight} Unity-based VR projects developed by more recent versions (e.g., after 2020.x). It is worth noting that we construct Group 1 (with relatively older Unity versions) primarily for a fair comparison with \textit{VRGuide}, as it only supports the ``click'' interaction. Compared with Group 1, Group 2 contains VR projects developed by recent Unity versions, which can support more diverse interactions (while not being supported in \textit{VRGuide}).

% Among the seven projects evaluated and open-source by , we
% However, the other eight projects are based on more recent versions that include a variety of interactions not supported by \textit{VRGuide}. Therefore, their Interactable Objects Coverage and Convergence Time Cost are not statistically meaningful. We thus divided the eleven projects into two groups to separately address efficiency (RQ1.1, only Group1) and coverage performance (RQ1.2, Group1 and Group2).

The \hn{11} projects selected for experiments represent a diverse and comprehensive subset of VR applications. 
% 种类
 Firstly, the chosen projects cover all of the most commonly featured VR application types as defined in~\cite{Virtual_Reality_Genres_Comparing_Preferences_in_Immersive_Experiences_and_Games}: 
(i) Action and Shooter (\texttt{\small VGuns}), (ii) Simulation (\texttt{\small VR-Basisc}, \texttt{\small VR-Room}, \texttt{\small UnityVR}), (iii) Adventure (\texttt{\small unity-vr-maze}, \texttt{\small VR-Adventure}), (iv) Puzzle (\texttt{\small Edutainment-Escape-Room}, \texttt{\small EscapeGameVR}), \hn{(v) Medical Care (\texttt{\small Parkinson-VR}), and (vi) Strategy Board Game (\texttt{\small VRChess})}. This diversity ensures that the experimental results are generalizable across different VR genres and interaction manners. 
Second, these projects cover a wide range of Unity versions, from older releases like 2019.4.2f1 (\texttt{\small UnityCityView}) to more recent versions such as 2022.3.7f1 (\texttt{\small VR-Adventure}), as well as the legacy version 5.4.1f1 (developed for \texttt{\small unity-vr-maze}). This version diversity ensures that our testing approach is evaluated across different Unity engine environments, demonstrating its robustness and compatibility. Moreover, the selected projects exhibit a variety of scales and complexities. For instance, the number of C\# scripts ranges from as few as 11 in \texttt{\small VR-Adventure} to over 275 in \texttt{\small Parkinson-VR}, with LOC spanning from around 260 to over 38,437.
This selection allows us to assess the testing tools' versatility from small and medium-sized VR applications to large ones. 
These projects also differ in the number of scenes and the number of GameObjects (GOs), with some projects like \texttt{\small EscapeGameVR} having 44 scenes, reflecting complex and rich environments, while others like \texttt{\small VR-Room} have fewer scenes but have potentially dense and complicated interactions. 
\textbf{Results on Group 1.}
%\end{itemize}
Fig.~\ref{fig:Results of RQ2 on Group1} plots EC versus time when testing all the projects in Group 1 where initial EC is identical across projects as initialization code (e.g., %\texttt{\small Start()} and 
        \texttt{\small Awake()}) runs immediately, regardless of interactions.
%\texttt{\small UnityCityView}, \texttt{\small UnityVR} and \texttt{\small unity-vr-maze}. 
%We observe from Figure~\ref{fig:Results of RQ2 on Group1}(\subref{fig:Coverage Performance of unity-vr-maze}) that \textit{VRExplorer} takes less time to reach convergence when it has higher EC and IOC than \textit{VRGuide}. We have similar observations from Figure~\ref{fig:Results of RQ2 on Group1}(\subref{fig:Coverage Performance of UnityCityView}) and Figure~\ref{fig:Results of RQ2 on Group1}(\subref{fig:Coverage Performance of UnityVR}). 
Table~\ref{tab:Results of RQ2 on Group1} shows the experimental results of \textit{VRExplorer} compared with baseline \textit{VRGuide} in Group 1. 
To quantitatively evaluate the performance improvement of \textit{VRExplorer} over \textit{VRGuide} (or other compared methods in \S~\ref{subsec:ablation}), \hn{we define the \textit{performance gain} of method $A$ over method $B$ in metric $M$ as follows,
\begin{equation}
  \vspace{-0.15cm}
\small
\label{eq:perf_gain}
    G_{AB}(M)=\frac{P_A(M)-P_B(M)}{P_B(M)}\times 100\%,
\end{equation} 
where $P_A(M)$ and $P_B(M)$ denote the performance of methods $A$ and $B$ with respect to metric $M$, respectively. For example, we evaluate the performance gain of \textit{VRExplorer} ($E$) over \textit{VRGuide} ($G$) in project \texttt{\small unity-vr-maze} in terms of EC as
$
G_{EG}(\text{EC})=\frac{P_E(\text{EC})-P_G(\text{EC})}{P_G(\text{EC})}\times 100\% = \frac{81.67 - 66.53}{66.53}\times 100\% = 22.8\%.
$
Similarly, $G_{EG}(\text{MC})$ and $G_{EG}(\text{IOC})$ are 16.7\% and 6.1\%, respectively, while the convergence time cost is reduced by 43.9\%.
For project \texttt{\small UnityCityView}, we have $G_{EG}(\text{EC})=36.3\%$, $G_{EG}(\text{MC})=27.6\%$, and $G_{EG}(\text{IOC})=66.7\%$.
%Our \textit{VRExplorer} also achieves consistent improvements in 
For project \texttt{\small UnityVR}, we have $G_{EG}(\text{EC})=17.1\%$, 
$G_{EG}(\text{MC})=9.1\%$, and a reduced convergence time cost of 12.5\%.  
Notably, \textit{VRExplorer} reaches 100\% IOC in all three projects, while \textit{VRGuide} achieves 100\% IOC only in \texttt{\small UnityVR}.  
In summary, these results demonstrate that \textit{VRExplorer} consistently outperforms \textit{VRGuide} in coverage and efficiency.}

\begin{table}[h] 
\caption{\hn{Results on Projects of Group 2}}
\centering
\vspace{-0.3cm}
\resizebox{0.9\linewidth}{!}{
\renewcommand{\arraystretch}{0.8}
\begin{tabular}{cccc}
\toprule
\textbf{Projects} & \textbf{Approaches} & \textbf{EC (\%)} & \textbf{MC (\%)} \\ \midrule
\multirow{2}{*}{\texttt{VR-Basics}} & \textit{VRGuide} & 41.38 & 53.22 \\
 & \textit{VRExplorer} & \colorbox{lightergray}{\parbox[c][1.0mm][c]{2cm}{\centering 80.17 (+93.8\%)}} & \colorbox{lightergray}{\parbox[c][1.0mm][c]{2cm}{\centering 91.93 (+72.8\%)}} \\ \midrule
\multirow{2}{*}{\texttt{VR-Room}} & \textit{VRGuide} & 40.97 & 50.63 \\
 & \textit{VRExplorer} & \colorbox{lightergray}{\parbox[c][1.0mm][c]{2cm}{\centering 77.61 (+89.4\%)}} & \colorbox{lightergray}{\parbox[c][1.0mm][c]{2cm}{\centering 83.54 (+65.0\%)}} \\ \midrule
\multirow{2}{*}{\texttt{VGuns}} & \textit{VRGuide} & 28.68 & 38.89 \\
 & \textit{VRExplorer} & \colorbox{lightergray}{\parbox[c][1.0mm][c]{2cm}{\centering 77.57 (+170.7\%)}} & \colorbox{lightergray}{\parbox[c][1.0mm][c]{2cm}{\centering 77.78 (+100.0\%)}} \\ \midrule
\multirow{2}{*}{\texttt{VR-Adventure}} & \textit{VRGuide} & 54.12 & 65.00 \\
 & \textit{VRExplorer} & \colorbox{lightergray}{\parbox[c][1.0mm][c]{2cm}{\centering 91.76 (+69.6\%)}} & \colorbox{lightergray}{\parbox[c][1.0mm][c]{2cm}{\centering 95.00 (+46.2\%)}} \\ \midrule
\multirow{2}{*}{\texttt{EE-Room}} & \textit{VRGuide} & 38.08 & 58.06 \\
 & \textit{VRExplorer} & \colorbox{lightergray}{\parbox[c][1.0mm][c]{2cm}{\centering 70.61 (+85.5\%)}} & \colorbox{lightergray}{\parbox[c][1.0mm][c]{2cm}{\centering 88.17 (+51.8\%)}} \\ \midrule
\multirow{2}{*}{\texttt{EscapeGameVR}} & \textit{VRGuide} & 41.77 & 55.26 \\
 & \textit{VRExplorer} & \colorbox{lightergray}{\parbox[c][1.0mm][c]{2cm}{\centering71.08 (+70.2\%)}} & \colorbox{lightergray}{\parbox[c][1.0mm][c]{2cm}{\centering 73.68 (+33.3\%)}} \\ \midrule
 \multirow{2}{*}{\hn{\texttt{Parkinson-VR}}} & \textit{VRGuide} & \hn{42.03} & \hn{53.85} \\
 & \textit{VRExplorer} & \colorbox{lightergray}{\parbox[c][1.0mm][c]{2cm}{\centering \hn{95.65 (+127.6\%)}}} & \colorbox{lightergray}{\parbox[c][1.0mm][c]{2cm}{\centering \hn{100.00 (+85.7\%)}}} \\ \midrule
 \multirow{2}{*}{\hn{\texttt{VRChess}}} & \textit{VRGuide} & \hn{10.74} & \hn{50.88} \\
 & \textit{VRExplorer} & \colorbox{lightergray}{\parbox[c][1.0mm][c]{2cm}{\hn{\centering 71.74 (\textbf{+568.0\%})}}} & \colorbox{lightergray}{\parbox[c][1.0mm][c]{2cm}{\hn{\centering 87.72 (+72.4\%)}}} \\ 
 \bottomrule
\end{tabular}
}
\label{tab:Result of RQ2 on Group2}
\vspace{-0.35cm}
\end{table}

%\begin{itemize}[leftmargin=0pt,itemindent=*] % 移除左侧缩进
%\item
\textbf{Results on Projects of Group 2.}
%\end{itemize}
Table~\ref{tab:Result of RQ2 on Group2} reports the results of \textit{VRExplorer} compared with \textit{VRGuide} on the \hn{eight} representative projects in Group 2 in EC and MC. Notably, we omit the IOC here mainly because \textit{VRGuide} exhibited lower IOC than our \textit{VRExplorer} due to its sole ``clicking'' interaction. 
 %During the experiment, we observed that on these more general subject projects compared to those in RQ1, \textit{VRGuide} exhibited very limited IOC due to its reliance solely on clicking. As a result, discussing IOC became meaningless. 
Therefore, we primarily focus on the code coverage (ELOC and MC) of the core development logic of VR projects.
%exclude the IOC and convergence time cost metrics from our analysis

We observe that our \textit{VRExplorer} consistently outperforms \textit{VRGuide} in all \hn{eight} projects. \hn{Notably, \textit{VRExplorer} achieves $G_{EG}(\text{EC}) = 567.97\%$ in \texttt{\small VRChess} significantly higher than \textit{VRGuide}. This is because this project contains numerous \textit{if-else} branches, difficult for \textit{VRGuide} to fully cover. In contrast, \textit{VRExplorer} can effectively explore most of these branches.}
%play chess like a real player, making it significantly easier to cover more branches.

%In particular, for project \texttt{\small VR-Basics}, \textit{VRExplorer} achieves EC and MC gains over \textit{VRGuide} by 93.8\% and 72.8\%, respectively. For project \texttt{\small VR-Room}, \textit{VRExplorer} achieves EC and MC gains by 89.4\% and 65.0\%, respectively. It is worth noting that \textit{VRExplorer} achieves EC and MC gains by 170.7\% and 100.0\%, respectively, in project \texttt{\small VGuns}, showcasing its strength in handling more complex and diverse VR testing scenarios.
%For \texttt{\small VR-Adventure}, EC gains by 69.6\%, and MC by 46.2\%. For \texttt{\small Edutain\\ment-Escape-Room}, EC gains by 85.5\%, and MC by 51.8\%. For \texttt{\small EscapeGameVR}, EC gains by 70.2\%, and MC by 33.3\%.  

Comprehensively considering experimental results in all \hn{11} projects in both Group 1 and Group 2, \textit{VRExplorer} has achieved an average EC gain of \hn{122.8\%} 
%= \((93.8 + 89.4 + 170.7 + 69.6 + 85.5 + 70.2 + 22.8 + 36.3 + 17.1 + 127.6 + 568.0 ) / 11 \) 
and average MC gain of \hn{52.8\%} 
%= \((72.8 + 65.0 + 100.0 + 46.2 + 51.8 + 33.3 + 16.7 + 27.6 + 9.1 + 85.7 + 72.4) / 9\) 
compared to \textit{VRGuide}. 
Experimental results demonstrate that \textit{VRExplorer} outperforms \textit{VRGuide} (in diverse performance metrics) consistently across various VR projects (even those developed with different Unity versions). 
%Compared to \textit{VRGuide}, \textit{VRExplorer} significantly outperforms in terms of code coverage for both ELOC and MC. In particular, the improvement in EC is remarkable across all projects, with the largest improvement seen in \texttt{\small VGuns} (170.7\%), highlighting the tool's ability to handle more complex and diverse VR testing scenarios. Although \textit{VRExplorer} achieves notable coverage improvements, a few projects still exhibit modest gains due to intrinsic design issues, which lead to disconnected NavMesh and unreachable objects.

%Although \textit{VRExplorer} achieves notable coverage improvements, a few projects still show relatively modest gains. These cases are primarily due to the intrinsic logic of the tested projects. For example, we find some inappropriate scene designs and gameobjects placement, which leads to disconnected NavMesh and unreachable objects. 

% For the projects with more generic subject matter, such as \texttt{\small VR-Basics} and \texttt{\small VR-Room}, \textit{VRExplorer} still outperforms \textit{VRGuide}, achieving substantial improvements in EC and MC. Even in cases where IOC is less relevant due to the limitations of \textit{VRGuide}, \textit{VRExplorer} maintains strong performance in code coverage. The overall trend indicates that \textit{VRExplorer} is capable of scaling effectively across different VR projects, regardless of the specific development packages or plugins used.

\noindent\fbox{\begin{minipage}{0.97\linewidth}
\textbf{Answer to RQ1:} Experimental results on all \hn{11} VR projects demonstrate that \textit{VRExplorer} outperforms the SOTA approach \textit{VRGuide} in EC, MC, and IOC. In particular, \textit{VRExplorer} achieves the average performance gains over \textit{VRGuide} in EC and MC by \hn{122.8\% and 52.8\%}, respectively, across all the projects. 
Moreover, \textit{VRExplorer} converges faster or comparably faster than \textit{VRGuide} while maintaining substantially higher coverage. 
Performance improvements are observed across diverse VR scenarios, demonstrating \textit{VRExplorer}'s strong generalizability in covering code and interactable VR objects during automated testing.
\end{minipage}}

%better efficiency than \textit{VRGuide} by achieving faster or comparably fast convergence while maintaining substantially higher coverage. This highlights \textit{VRExplorer}'s ability to balance convergence time with thorough testing.

%\textbf{\textit{Answer to RQ1.1 (Efficiency Comparison):}}
%Results on Group1 projects demonstrate that \textit{VRExplorer} performs better efficiency than \textit{VRGuide} by achieving faster or comparably fast convergence while maintaining substantially higher coverage. This highlights \textit{VRExplorer}'s ability to balance convergence time with thorough testing.

% \textbf{\textit{Answer to RQ1.2 (Coverage Performance Comparison):}}
% Results on Group1 and Group2 projects demonstrate that \textit{VRExplorer} consistently achieves significantly higher coverage than \textit{VRGuide} across all eleven evaluated projects. These improvements hold across diverse real-world VR scenarios, demonstrating strong generalizability and versatility in covering code and interactable objects during automated testing.
% \textit{VRExplorer} demonstrates excellent generalizability and versatility when applied to a variety of VR projects developed with different VR ecosystem packages and plugins. It consistently outperforms \textit{VRGuide} in terms of ELOC and MC across all evaluated projects. These results suggest that \textit{VRExplorer} is a robust tool for testing VR applications in diverse development environments, making it suitable for a wide range of VR testing scenarios.
% \end{tcolorbox}
% \vspace{-0.3cm}

\subsection{Ablation Study}\label{subsec:ablation}

To assess the contribution of different components, we perform an ablation study by selectively removing modules from the proposed \textit{VRExplorer}. 
\hn{Given that \textit{VRExplorer} is developed on top of the EAT framework, which decomposes VR interactions into reusable action units. Thus, EAT serves as the core exploration task of inherently focusing on interacting with objects. As a result, our ablation primarily targets the interaction-related components within EAT. 
The EAT framework constitutes the key methodological contribution of \textit{VRExplorer} and is the only component that can be meaningfully isolated while preserving system functionality. 
In contrast, other modules and components serve as indispensable prerequisites for enabling exploration in VR scenes and are tightly integrated into the overall pipeline, rendering their removal infeasible for independent evaluation.
}

We then evaluate these methods by comparing them with the full-fledged \textit{VRExplorer} framework in terms of EC and MC. Notably, we also compare them with \textit{VRGuide} to investigate the performance improvement contributed by each module. 

\begin{table}[t]
\centering
\caption{Results of Ablation Study}
     \vspace{-0.2cm}
\resizebox{0.91\linewidth}{!}{
\renewcommand{\arraystretch}{0.8}
\begin{tabular}{cccc}
\toprule
\textbf{Projects} & \textbf{Approaches} & \textbf{EC (\%)} & \textbf{MC (\%)} \\ 
\midrule
\multirow{4}{*}{\texttt{VR-Basics}} & \textit{VRGuide} & \multicolumn{1}{c}{41.38} & 53.22 \\ 
 & \textit{VRExplorer} & \textbf{\underline{80.17}} & \textbf{\underline{91.93}} \\ 
 & \textit{VRExplorer} w/o \textit{T} & \colorbox{lightergray}{\parbox[c][1.0mm][c]{2cm}{\centering 68.10 (-15.0\%)}} & \colorbox{lightergray}{\parbox[c][1.0mm][c]{2cm}{\centering 77.42 (-15.9\%)}} \\ 
 & \textit{VRExplorer} w/o \textit{Tf} & \colorbox{lightergray}{\parbox[c][1.0mm][c]{2cm}{\centering 59.24 (-26.1\%)}} & \colorbox{lightergray}{\parbox[c][1.0mm][c]{2cm}{\centering 70.00 (-16.2\%)}} \\ 
 
 \midrule
 
\multirow{4}{*}{\texttt{VR-Room}} & \textit{VRGuide} & \multicolumn{1}{c}{40.97} & 50.63 \\ 
 & \textit{VRExplorer} & \textbf{\underline{77.61}} & \textbf{\underline{83.54}} \\ 
 & \textit{VRExplorer} w/o \textit{G} & \colorbox{lightergray}{\parbox[c][1.0mm][c]{2cm}{\centering 58.52 (-24.6\%)}} & \colorbox{lightergray}{\parbox[c][1.0mm][c]{2cm}{\centering 69.62 (-16.4\%)}} \\ 
 & \textit{VRExplorer} w/o \textit{T} & \colorbox{lightergray}{\parbox[c][1.0mm][c]{2cm}{\centering 64.12 (-17.3\%)}} & \colorbox{lightergray}{\parbox[c][1.0mm][c]{2cm}{\centering 67.00 (-19.7\%)}} \\ 

\midrule

 \multirow{4}{*}{\texttt{VGuns}} & \textit{VRGuide} & \multicolumn{1}{c}{28.68} & 38.89 \\ 
 & \textit{VRExplorer} & \textbf{\underline{77.57}} & \textbf{\underline{77.78}} \\ 
 & \textit{VRExplorer} w/o \textit{TG} & \colorbox{lightergray}{\parbox[c][1.0mm][c]{2cm}{\centering 50.37 (-35.3\%)}} & \colorbox{lightergray}{\parbox[c][1.0mm][c]{2cm}{\centering 61.11 (-16.7\%)}} \\ 
 & \textit{VRExplorer} w/o \textit{AE} & \colorbox{lightergray}{\parbox[c][1.0mm][c]{2cm}{\centering 65.07 (-16.1\%)}} & \colorbox{lightergray}{\parbox[c][1.0mm][c]{2cm}{\centering 63.89 (-17.9\%)}} \\ 
 
 \bottomrule
\end{tabular}}
\label{tab:RQ3}
\smallskip
    % \footnotesize \textbf{\textit{Notes:}} \textit{T}, \textit{Tf}, and \textit{G} represent \textit{Triggerable}, \textit{Transformable}, and \textit{Grabbable}, respectively. "w/o" denotes "without."
         \vspace{-0.5cm}
\end{table}

% \begin{itemize}[leftmargin=0pt,itemindent=*] % 移除左侧缩进
% \item 
\textbf{Ablation Experiment Configuration.}
As described in \S~\ref{subsection: EAT Framework}, interaction behaviors play a crucial role in the EAT framework. To investigate these interaction behaviors, we remove a specific interaction from a tested project's modules (as this module is used in this project). 
For example, we obtain VRExplorer without (w/o) \textit{T} by removing the \textit{Triggerable} module from the Entity layer's interface \texttt{\small Triggerable}, the Action layer's class \texttt{\small TriggerAction}, all tasks involve triggering in the Task layer, and the corresponding Mono C\# scripts \texttt{\small XRTriggerable.cs}. Similarly, we obtain \textit{VRExplorer}'s other ablated variants: w/o \textit{Tf} (without \textit{Transformable}), w/o \textit{G} (without \textit{Grabbable}), w/o \textit{TG} (without both \textit{Triggerable} and \textit{Grabbable}), and w/o \textit{AE} (without \textit{autonomous event}).
%Modules in our approach represent the corresponding part of a specific behavior in the EAT framework's three layers. In RQ1, we use the full module \textit{VRExplorer}, while in RQ2, to explore the effect of different modules of the EAT framework on its coverage performance, we
% \end{itemize}
% \begin{table}[h]
% \caption{Results of RQ3}
% \resizebox{\linewidth}{!}{
% \begin{tabular}{|c|c|cc|}
% \hline
% \multirow{2}{*}{\textbf{Project Name}} & \multirow{2}{*}{\textbf{Approach}} & \multicolumn{2}{c|}{\textbf{Metrics}} \\ \cline{3-4}
%  &  & \multicolumn{1}{c|}{\textbf{\begin{tabular}[c]{@{}c@{}}ELOC \\ Coverage\\ (\%)\end{tabular}}} & \textbf{\begin{tabular}[c]{@{}c@{}}Method \\ Coverage\\ (\%)\end{tabular}} \\ \hline
% \multirow{4}{*}{\textbf{VR-Basics}} & VRGuide & \multicolumn{1}{c|}{41.38} & 53.22 \\ \cline{2-4}
%  & VRExplorer & \multicolumn{1}{c|}{80.17} & 91.93 \\ \cline{2-4}
%  & VRExplorer (remove Triggerable) & \multicolumn{1}{c|}{68.10 (-15.0\%)} & 77.42 (-15.9\%) \\ \cline{2-4}
%  & VRExplorer (remove Transformable) & \multicolumn{1}{c|}{59.24 (-26.1\%)} & 70.00 (-16.2\%) \\ \hline
% \multirow{4}{*}{\textbf{VR-Room}} & VRGuide & \multicolumn{1}{c|}{40.97} & 50.63 \\ \cline{2-4}
%  & VRExplorer & \multicolumn{1}{c|}{77.61} & 83.54 \\ \cline{2-4}
%  & VRExplorer (remove Grabbable) & \multicolumn{1}{c|}{58.52 (-24.6\%)} & 69.62 (-16.4\%) \\ \cline{2-4}
%  & VRExplorer (remove Triggerable) & \multicolumn{1}{c|}{64.12 (-17.3\%)} & 67.00 (-19.7\%) \\ \hline
% \end{tabular}
% }
% \label{tab:RQ3}
% \end{table}

% \begin{itemize}[leftmargin=0pt,itemindent=*] % 移除左侧缩进
% \item 
\textbf{Results of Ablation Study.}
% \end{itemize}
Table~\ref{tab:RQ3} presents the experimental results of the ablation study in three projects: \texttt{\small VR-Basics}, \texttt{\small VR-Room}, and \texttt{\small VGuns}, where the best results are highlighted with underline and bold.
These ablation experiments cover all the \textit{VRExplorer}'s variants and span all the Unity versions of the projects in Group 2 (from 2020 to 2022). Specifically, we compare \textit{VRGuide} and \textit{VRExplorer} with its five ablated variants: \textit{VRExplorer} w/o \textit{G}, w/o \textit{T}, w/o \textit{Tf}, w/o \textit{TG}, and w/o \textit{AE}. To quantitatively evaluate the effect of removing each module, we also calculate the performance gain of a method over another one according to Eq.~\eqref{eq:perf_gain}.

%      \vspace{-0.35cm}
% \begin{equation}
% \text{Relative Variation} = \frac{\text{Metric}_{\text{Ablated}} - \text{Metric}_{\text{Full}}}{\text{Metric}_{\text{Full}}} \times 100\%
% \label{eq:ablation}
% \end{equation}

In \texttt{\small VR-Basics}, the full-fledged \textit{VRExplorer} achieves the best performance in terms of 80.17\% EC and 91.93\% MC. We observe that removing the \textit{Triggerable} module causes a 15.0\% ($\frac{68.10 - 80.17}{80.17} \times 100\%$) decrease in EC and a 15.9\% decrease in MC and the removal of the \textit{Transformable} module leads to a 26.1\% decrease in EC and 16.2\% in MC. 
%These results indicate that both \textit{Triggerable} and \textit{Grabbable} interaction modules contribute significantly to \textit{VRExplorer}'s performance.
%, with \textit{Transformable} playing a slightly smaller role.
In \texttt{\small VR-Room}, we also find that the removal of the \textit{Grabbable} module has the most pronounced effect, i.e., reducing EC by 24.6\% and MC by 16.4\%. Disabling the \textit{Triggerable} module also leads to an ELOC decrease of 17.3\% and a MC decrease of 19.7\%. 
%These results confirm the crucial contribution of each interaction module to \textit{VRExplorer} despite various portions across different projects.
%, depending on the interactive design.
In \texttt{\small VGuns},  we observe that the removal of both the \textit{Triggerable} and \textit{Grabbable} modules causes a 35.3\% decrease in EC and a 16.7\% decrease in MC. 
Moreover, the removal of the \textit{autonomous event} module causes a 16.1\% decrease in EC and a 17.9\% decrease in MC, indicating that the \textit{autonomous event} module as described in Step~\circlednumber{3} (\S~\ref{subsection: VRExplorer}) also contributes significantly. 
These results confirm the significant contribution of each interaction module to \textit{VRExplorer}.
%despite various portions across different projects.
%, depending on the interactive design

Overall, removing any individual module leads to a substantial drop in code coverage (in terms of EC and MC), further confirming the importance of integrating all three interaction-aware modules in the EAT framework. The full-fledged \textit{VRExplorer} consistently yields the highest code coverage.

%\vspace{-0.15cm}

% \begin{tcolorbox}
% [width=\linewidth, colframe=black!50, colback=black!5, coltitle=black, boxrule=1pt, arc=4pt, breakable]
\noindent\fbox{
\begin{minipage}{0.97\linewidth}
\textbf{{Answer to RQ2:}}  
Removing any module from the EAT leads to a substantial decrease in EC and MC. The degree of impact caused by the ablated module varies across diverse projects and interaction types. %, while \textit{Triggerable} and \textit{Grabbable} modules are observed to have the most significant influence.
These results validate the necessity and effectiveness of integrating all the modules in \textit{VRExplorer} for comprehensive interaction-aware VR testing.
\end{minipage}}
% \end{tcolorbox}
% \vspace{-0.3cm}

\subsection{Bug Detection in VR Projects}

% 说明bug检测结果
Since the primary goal of software testing is to detect or identify potential bugs, we also investigate whether the proposed \textit{VRExplorer} can detect real-world bugs in VR projects. We have scanned all \hn{11} projects through comprehensive testing conducted by \textit{VRExplorer}. Thereafter, we have successfully detected two \textit{functional}\footnote{A functional bug refers to a bug that causes the VR application not to work as expected, while a non-functional bug denotes a bug not directly related to the functionality (may be relevant to performance and usability issues).} bugs and one \textit{non-functional} bug in total,
%\footnote{%Detailed bug report:\url{https://anonymous.4open.science/r/VRExplorer-683A/Artifacts/Bug_Report.md}.}
from projects \texttt{\small EscapeGameVR}, \texttt{\small UnityCityView}, and \texttt{\small EscapeGameVR}. 
%Herein, a functional bug refers to a bug that causes the VR application not to work as expected, while a non-functional bug denotes a bug not directly related to the functionality (may be relevant to performance and usability issues). 
Notably, two functional bugs have been found in \texttt{\small EscapeGameVR} and \texttt{\small UnityCityView} while we have detected one non-functional bug in \texttt{\small EscapeGameVR}. Moreover, the bug from \texttt{\small UnityCityView} has been fixed by developers (after checking the commits), while two bugs from \texttt{\small EscapeGameVR} reported by us have not been confirmed by developers so far.
Our \textit{VRExplorer} can detect all these VR bugs, which are either functional or non-functional. Although \textit{VRGuide} is able to detect the bug in project \texttt{\small UnityCityView}, it misses the other two bugs in \texttt{\small EscapeGameVR}.
%Among these three bugs, our \textit{VRExplorer} is able to detect and report all of the functional bugs and can also detect a non-functional bug after a simple manual check of the reported information. While \textit{VRGuide} is able to detect the bug in \texttt{\small UnityCityView}, but misses the two bugs in \texttt{\small EscapeGameVR}.

\hn{
\textbf{Detected Bugs Validity Confirmation.} To further confirm the validity of detected bugs, we validated all reported issues through Unity console logs, runtime exception traces, and script-trigger chains. Among them, the bug in \texttt{\small UnityCityView} has been confirmed and fixed by its developers (evidenced by later commits). For the two bugs in \texttt{\small EscapeGameVR}, we performed root-cause analyses: the functional bug is an \textit{\small Unassigned Reference Exception}, consistent with CWE-395~\cite{cwe-395} and CodeQL’s rule on improper initialization~\cite{codeql-nullref}, while the non-functional bug is caused by a missing \texttt{\small ArrowPrefab} resource, which is a common Unity issue documented in the official Prefab guidelines~\cite{unity-brokenprefabasset}. We also reported these issues to the original developers of \texttt{\small EscapeGameVR}, but no response was received till now. 
}

\begin{figure}[t] 
    \centering
    \includegraphics[width=0.9\linewidth]{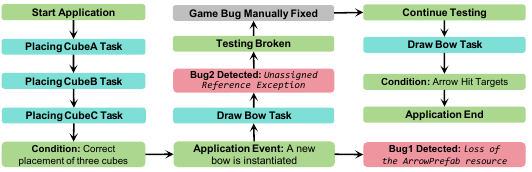}  
        \vspace{-0.15cm}
    \caption{\hn{Testing Process of \texttt{\small EscapeGameVR}}}
    \label{fig:Testing Process of EscapeGameVR}
       \vspace*{-0.3cm}
\end{figure}

\begin{figure}[t] 
    \centering
    \includegraphics[width=0.8\linewidth]{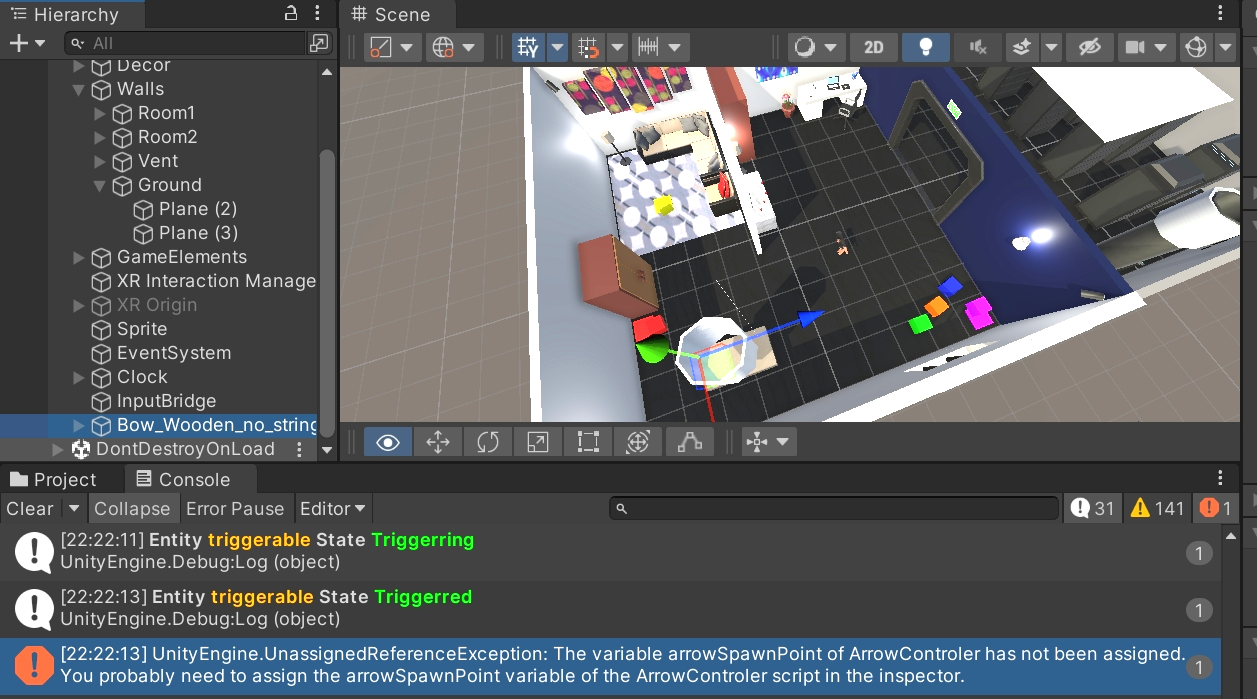}  
        \vspace{-0.15cm}
    \caption{A Detected Bug in \texttt{\small EscapeGameVR}}
    \label{fig:A Detected Bug in EscapeGameVR}
       \vspace*{-0.3cm}
\end{figure}

%介绍funcional bug
\textbf{Functional Bug Root-cause Analysis.} Fig.~\ref{fig:Testing Process of EscapeGameVR} shows the testing process in Unity and Fig.~\ref{fig:A Detected Bug in EscapeGameVR} shows a screenshot of one functional bug detected by our \textit{VRExplorer} in project \texttt{\small EscapeGameVR}. 
This bug occurs due to \textit{assignment exception}, which can be explained as follows. When a bow releases an arrow, the method \texttt{\small ReleaseArrow()} is invoked with the attempt to access the \texttt{\small arrowSpawnPoint} field of the \texttt{\small ArrowController} class. However, the developers have mistakenly failed to assign a value to this field (or the assignment was unintentionally lost), thereby resulting in an \textit{\small Unassigned Reference Exception}~\cite{cwe-395, codeql-nullref} bug, which is a Unity-specific runtime exception occurring when a serialized reference field (typically declared as \texttt{\small public} or marked with \texttt{\small [SerializeField]}~\cite{unity-serializefield}) in a \texttt{\small MonoBehaviour} class is accessed without being assigned a value in the Unity Inspector. 

%介绍另一个non-functional bug
\textbf{Non-functional Bugs Root-cause Analysis.} Another non-functional bug in \texttt{\small EscapeGameVR} is caused by the loss of the \texttt{\small ArrowPrefab} resource. This bug can be identified by our \textit{VRExplorer} through a simple manual inspection. %(Although \textit{VRExplorer} does not report it directly, the bug can be identified through a simple manual inspection after the functional bug we described above.) 
The reason why these two bugs are not detected by \textit{VRGuide} is that triggering these two bugs requires complicated actions and multiple types of interactions, while \textit{VRGuide} cannot handle them.
% 具体来说，任务为什么复杂，以及VRGuide为什么不行，以及我们的工具为什么行
Differently, our \textit{VRExplorer} can complete this difficult testing task through our model-based framework. In particular, \textit{VRExplorer} requires the correct placement of three cubes onto platforms of the corresponding color, thereby ensuring that none of them fall outside the designated platform areas. Once this condition is satisfied, a new bow is instantiated. Consequently, triggering the bug needs the bow to be drawn and released by \textit{VRExplorer}. This type of complex interactive task contains two or even more interactive patterns% (e.g., \textit{Transformable}, \textit{Grabbable}, and \textit{Triggerable})
, which can not be completed by \textit{VRGuide} (only supporting ``clicking'' interaction), while our approach handles this task properly.

% The root-cause analysis and detected bug validity confirmation demonstrate that our tool can not only detect real-world bugs but also reveal complex VR-specific issues that may remain latent in open-source projects.

%through our proposed model-based framework.
%\vspace{-0.05cm}
% \begin{tcolorbox}
% [width=\linewidth, colframe=black!50, colback=black!5, coltitle=black, boxrule=1pt, arc=4pt, breakable]
\noindent\fbox{\begin{minipage}{0.97\linewidth}
\textbf{Answer to RQ3:}  
%Testing VR projects, 
\textit{VRExplorer} can successfully detect three real-world bugs in VR projects. 
%Attributed to the generalizability of the EAT framework, 
More importantly, our \textit{VRExplorer} is capable of detecting complex VR bugs, which can only be triggered after complex interactions.
\end{minipage}}
%\end{tcolorbox}

\section{Discussions}
\label{sec: Discussion}
%\begin{itemize}[leftmargin=0pt,itemindent=*] % 移除左侧缩进

%\item 
\textbf{Threats to External Validity}.
% \textbf{Threats to External Validity.} 
%The main external validity threat, 
Although we have covered as many types of VR applications (developed by diverse Unity versions) as possible, there are still types of scenarios and interaction patterns not fully covered in our approach. Moreover, while our framework is currently tailored for VR applications developed in Unity, it can be extended to support other engines, such as Unreal Engine~\cite{UE5}, with further adaptation, thereby enabling its broader applicability.

%\item 
\textbf{Threats to Internal Validity}.
% \textbf{Threats to Internal Validity.}
% 为了减少误差，确保有效性，在Core Code选择阶段，我们安排了4位测试人员分别单独选择后，再讨论得出最终结果。在接口实现阶段，我们让4位测试人员共同讨论，确保得到一致的定制化实现。
The primary threats to internal validity arise from potential human errors during the selection of Core Code (CC) and the implementation of the interface. Specifically, during the CC selection phase, errors could be introduced due to subjective judgment. To minimize this threat, we have leveraged a majority-voting mechanism, where four test engineers (with VR developing experience) have independently selected code segments. The final decision has been made according to the majority vote.
Moreover, during the interface implementation phase, potential inconsistencies or biases among test engineers are also present. To address this threat, we have arranged for four test engineers to collaboratively discuss the design and reach a consensus.
%thereby helping to mitigate discrepancies and ensure a consistent yet customized implementation.

\hn{
\textbf{Discussion on Path-finding Algorithms}.
%\label{subsec:Discussion on Path-finding Algorithms}
%Beyond interaction modeling, 
We also revisit the design choice of adopting the Greedy algorithm for navigation. 
As described in \S~\ref{subsection: VRExplorer}, \textit{VRExplorer} supports both a Greedy strategy and a Backtracking-with-Pruning (B\&P) algorithm. The latter solves the navigation task as a Traveling Salesman Problem (TSP) to compute globally optimal Hamiltonian paths. 
To quantify their differences, we performed additional experiments on randomized tasks with up to 100 interactable objects. 
Results show that the B\&P algorithm achieves a 17.5\% improvement in runtime efficiency (1233.70s vs. 1494.41s over 100 rounds) over the Greedy baseline. 
However, this improvement comes at the cost of significantly higher computational overhead, leading to reduced frame rates. 
This trade-off confirms our design decision: the Greedy algorithm remains more practical for maintaining smooth interactive performance in real-time testing than the B\&P algorithm.
%while the B\&P algorithm can provide globally optimal solutions
}

\begin{figure}[t] 
\vspace{-0.3cm}
    \centering
    \includegraphics[width=0.9\linewidth]{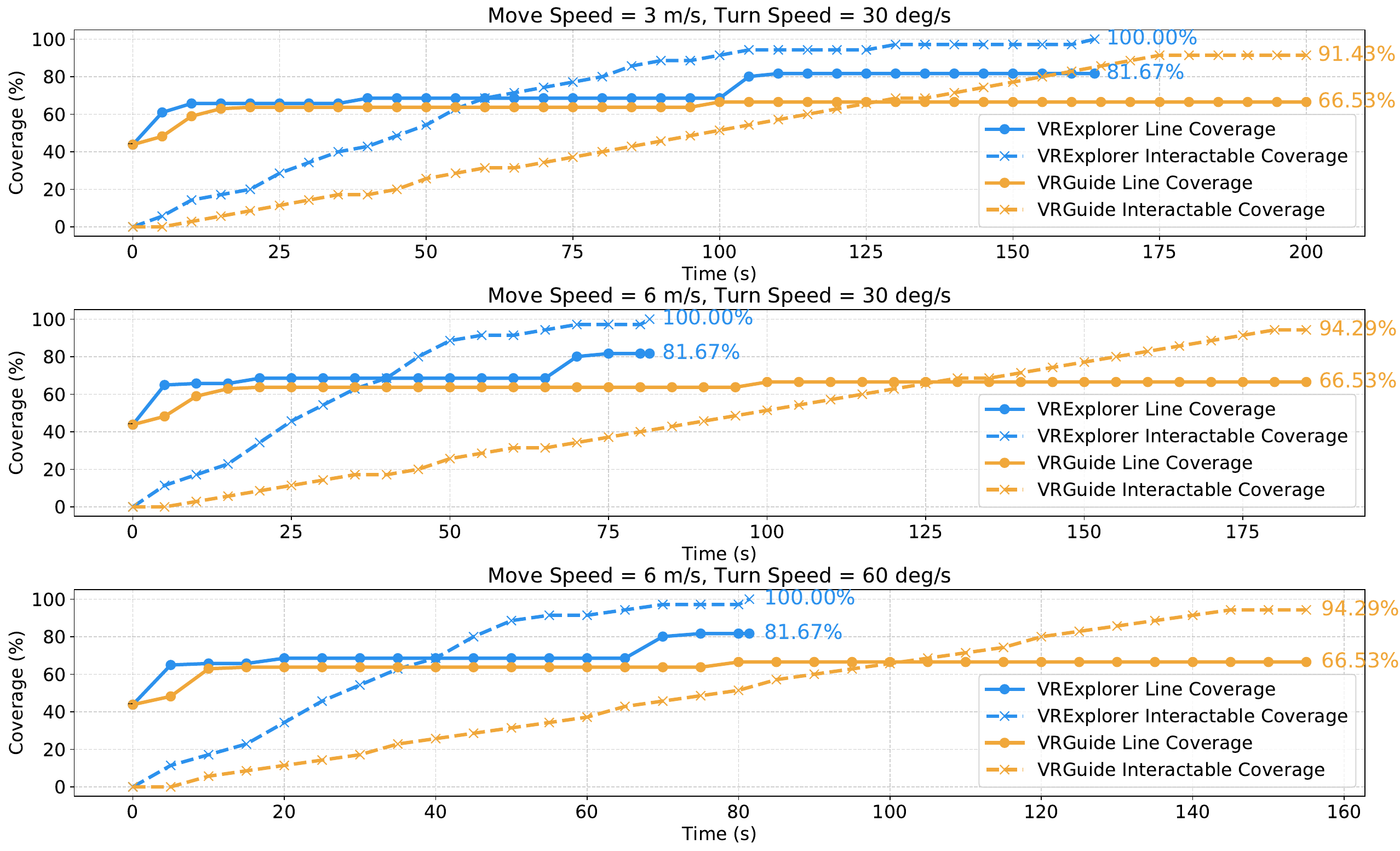}  
    \vspace{-0.3cm}
    \caption{Coverage on different MS and TS in \texttt{\small unity-vr-maze}}
    \label{fig:Coverage Performance of unity-vr-maze on different speed parameters}
    \vspace{-0.4cm}
\end{figure}

%\item
\textbf{Impacts of Speed Parameters}.
% \textbf{Impacts of Speed Parameters.} 
%\label{sec:Preliminary Experiment on Speed Parameters} 
To explore optimal speed parameters, we test three sets of values in \texttt{\small unity-vr-maze}. %, which has also been evaluated by \textit{VRGuide} in~\cite{VRGuide} and includes 7 core C\# scripts. 
Specifically, we have chosen (i) MS = 3 m/s, TS = 30 deg/s, (ii) MS = 6 m/s, TS = 30 deg/s; and (iii) MS = 6 m/s, TS = 60 deg/s. 
%During the experiments, all of the \textit{initial position} metric is fixed at \texttt{\small (0,4.5,0)} to ensure the test tool be legally standing on the floor.
Fig.~\ref{fig:Coverage Performance of unity-vr-maze on different speed parameters} shows the coverage performance of  \textit{VRExplorer} compared to \textit{VRGuide} in \texttt{\small unity-vr-maze} with different speed parameters. 
We observe the same trends for all three parameters. Notably, \textit{VRExplorer} takes less time to reach convergence than \textit{VRGuide} with higher EC and IOC. These results also indicate that TS has a minimal impact on efficiency, whereas MS shows a more significant effect. 

 % In Figure~\ref{fig:Coverage Performance of unity-vr-maze on different speed parameters}, what needs additional explanation is that the EC all starts from 43.82\%, which is due to the reason that all VR projects' code for initialization, such as \texttt{\small Start()} and \texttt{\small Awake()} will be immediately executed, which have no relationship with core interaction code.

%\item 

\hn{
\textbf{Generality of VRExplorer}.
% \label{subsec:Generality of VRExplorer}
% \ZZY{Upd: 新增一个段落说明VRExplorer的通用的底层逻辑（框架使得与具体输入解耦，与项目无关）。结合OOP来讲，本身面对通用泛型）}
Our approach systematically unifies these diverse interactions by decomposing VR objects into a finite set of interactable features, which are then represented as abstract actions. Based on OOP principles, the Entity Layer and Action Layer encapsulate input modalities and target objects separately, effectively treating them as reusable and generic components. Within this feature–action hierarchy, abstract actions are further integrated into PFSM as semantic nodes and transitions, allowing frequent and domain-specific behaviors to be expressed uniformly. This design ensures generality across projects while preserving extensibility to new interactions. For those not mentioned in \S~\ref{sec:Approach}, it only requires modeling a simple composition of task nodes and an additional configuration, without modifying the core logic. This OOP-based design allows the framework to scale naturally to a wide spectrum of VR interactions. 
For example, \textit{\small GUI Interactions} (e.g., typing and pressing) can be represented by customized tasks composed of \textit{\small Trigger Action} nodes, which provide event lists and function call chains for simulating different GUI interactions. 
%Game logic-related interactions, such as \textit{\small Shoot/Fire/Attack} and \textit{\small Dialogue}, can be modeled as sequences of \textit{Trigger Actions} linked to state transitions in PFSM, capturing both branching logic and user choices.
Similarly, \textit{\small Hand Gesture}-based interactions can be supported by treating gestures as input-trigger components in the Entity Layer, while a dedicated \textit{\small Gesture Action} class in the Action Layer encapsulates concrete gestures such as swipe, pinch, and point. These gestures can then be mapped to task transitions in PFSM, e.g., ``swipe-to-turn-the-page'' or ``pinch-to-zoom,'' in \S~\ref{subsection: VRExplorer}. In summary, the EAT framework is inherently \textit{agnostic} to input types: each project only needs to define its input actions and associated targets, after which abstraction and task composition are handled systematically through the generic OOP framework.
}

\textbf{Limitations and Future Work}.
% \textbf{Limitations and Future Work.} 
%At present, 
\textit{VRExplorer} still needs to analyze scenes manually and implement customized interfaces, inevitably introducing some extra workloads (despite being low) and a certain learning curve for test engineers. As future work, we will improve \textit{VRExplorer}'s capability of automatically understanding VR scenes inspired by recent advances in multi-modal large models. Further, we will also automate the interface implementation and task model generation.
%with the advent of large language models. 
%so that it can analyze scene visual screenshots, scene information (code and object association information), etc., to automatically realize interface implementation and task model generation.

%Our future works, on the one hand, we hope to add multi-modal big models into \textit{VRExplorer}, so that it can analyze scene visual screenshots, scene information (code and object association information), etc., to automatically realize interface implementation and task model generation.
%On the other hand, we will continue to do more experiments on other projects and try to detect more real-world bugs.

%\end{itemize}

%\subsection{Statistical Analysis of the Dataset and Quantitative Metrics of the Abstract Model Process}

\section{Related Work}

\label{sec:related-work}
%VR application testing shares several similarities with 3D game testing, as both are predominantly developed using Unity. 
%This section briefly reviews related work on automated game as well as mobile application testing, Reinforcement Learning (RL)-based, and automated VR application testing.

%\begin{itemize}[leftmargin=0pt,itemindent=*] % 移除左侧缩进
%\item 
\textbf{Automated Game Testing.} 
%Since many VR applications have been developed by game engines, there are several game testing approaches related to our work. 
As a code-based data augmentation technique, \textsc{Glib}~\cite{1GLIB_towards_automated_test_oracle_for_graphically_rich_applications} can automatically detect game GUI
glitches. 
Macklon et al.~\cite{Automatically_Detecting_Visual_Bugs_in_HTML5_Canvas_Games} present an approach to automatically detect visual bugs in \texttt{\small <canvas>} games.
Prasetya et al.~\cite{Navigation_and_Exploration_in_3D_Game_Automated_Play_Testing} leverage graph-based path-finding techniques in automated game navigation and exploration. As a Java-based multi-agent programming framework, \textsc{ix4XR}~\cite{An_agent_based_approach_to_automated_game_testing_an_experience_report, Using_an_Agent_Based_Approach_for_Robust_Automated_Testing_of_Computer_Games} facilitates game testing by enabling test agents to interact with the game under test.
%through an interface called \textit{Environment}. 
As a Belief-Desire-Intention library, \textsc{Aplib}~\cite{Aplib_Tactical_Agents_for_Testing_Computer_Games} supports developing intelligent agents capable of executing complex testing tasks. 
Ferdous et al.~\cite{Search_Based_Automated_Play_Testing_of_Computer_Games:_A_Model-Based_Approach} propose a model-based approach by leveraging EFSM to model game behavior. %and achieving high coverage and uncovering unknown faults in a 3D game. 
However, these game testing approaches are generally tailored to specific types of games instead of VR applications. 
%and cannot be adopted for VR testing.

%\item 
\textbf{Automated Mobile Application Testing.}
\textit{Stoat}~\cite{Guided_stochastic_model-based_GUI_testing_of_Android_apps} is a stochastic model-based testing tool for Android apps with combined dynamic and static analysis to generate tests. 
Chen et al. ~\cite{Model_based_GUI_Testing_For_HarmonyOS_Apps} propose a model-based GUI testing approach for HarmonyOS applications with the adoption of the \textit{arkxtest}\cite{Arkxtest} framework.
%for automation. %introducing a Page Transition Graph (PTG) constructed from the abstract syntax tree (AST) and call graph (CG), and utilizes the \textit{arkxtest}\cite{Arkxtest} framework for automation.
AutoConsis~\cite{AutoConsis} leverages a specially tailored Contrastive Language-Image Pre-training (CLIP) multi-modal model to automatically analyze Android 2D GUI pages.
\textsc{Fastbot2}~\cite{Fastbot2} leverages a probabilistic model that memorizes key information for testing based on a model-guided testing strategy. %is another automated model-based GUI testing tool, where the author proposes a probabilistic model that memorizes key information for testing, and a model-based guided testing strategy.
\textsc{Kea}~\cite{General_and_Practical_Property-based_Testing_for_Android_Apps} is a property-based testing tool for finding functional bugs in Android apps. 
However, these mobile application testing approaches can only address 2D GUI testing. % on Android apps. and are incapable of testing 3D VR applications. %, such as 3D scene exploration.

%\item 
\textbf{RL-Based Testing.}
%Prior research shows that RL enhances the efficiency of game testing.
Tufano et al.~\cite{Using_Reinforcement_Learning_for_Load_Testing_of_Video_Games} employ RL to train an agent to play games in a human-like manner to identify areas that lead to FPS drops.
\textsc{RLbT}~\cite{Towards_Agent_Based_Testing_of_3D_Games_using_Reinforcement_Learning} applies a curiosity-based RL approach to automate game testing by maximizing coverage. %Empirical evaluation applies \textsc{RLbT} to a 3D game called Lab Recruits and compares the curiosity-based approach with several alternative baselines.
Bergdahl et al.~\cite{Augmenting_Automated_Game_Testing_with_Deep_Reinforcement_Learning} adopts a modular approach, in which RL complements classical test scripting.
%rather than replacing it. %Their work expands to include exploit detection and continuous action simulation, such as mouse movements, with a primary focus on navigation rather than combat.
\textit{Wuji}~\cite{Wuji_Automatic_Online_Combat_Game_Testing_Using_Evolutionary_Deep_Reinforcement_Learning} leverages evolutionary algorithms, RL, and multi-objective optimization for game testing. %is a game testing framework that leverages evolutionary algorithms, RL, and multi-objective optimization. % to perform automated game testing.
However, %without our framework,
RL-based approaches alone are very limited in complex VR scene exploration.

%\item 
\textbf{VR Application Testing.}
Rzig et al.~\cite{Virtual_Reality_VR_Automated_Testing_in_the_Wild_A_Case_Study_on_Unity_Based_VR_Applications} analyze 314 open-source VR applications, revealing that 79\% of them lack automated tests. %~\citet{Virtual_Reality_VR_Automated_Testing_in_the_Wild_A_Case_Study_on_Unity_Based_VR_Applications} conducts an empirical study on 314 open-source VR applications. The analysis shows that 79\% of the evaluated VR projects do not include any automated tests. For the projects that do, the median ratio of functional methods to test methods is lower than that of other types of software projects.
Harms~\cite{Automated_Usability_Evaluation_of_Virtual_Reality_Applications} proposes an automated approach that extracts task trees from real VR usage recordings to detect usability smells without predefined tasks or settings.
%~\citet{Automated_Usability_Evaluation_of_Virtual_Reality_Applications} presents a fully automated VR application testing approach that does not require users to perform predefined tasks in fixed test settings. Instead, it analyzes recordings of actual usage sessions to generate task trees. These task trees are then examined to identify usability smells, which refer to patterns of user behavior that indicate potential usability issues.
\textsc{PredART}~\cite{PredART} predicts human ratings of virtual object placements, serving as test oracles in AR testing.
\textit{VRTest}~\cite{VRTest} extracts information from a VR scene and controls the user's camera to explore the scene and interact with virtual objects. % using specific testing strategies. %It includes two built-in strategies: VRMonkey, which uses pure random exploration, and VRGreed, which applies a greedy algorithm to explore interactable objects in VR scenes.
\textit{VRGuide}~\cite{VRGuide} applies a computational geometry technique called Cut Extension to optimize camera routes for covering all interactable objects. 
Qin and Weaver~\cite{Utilizing_Generative_AI_for_VR_Exploration_Testing_A_Case_Study} explore Generative AI for field of view analysis in VR testing. %, showing GPT-4o achieves 63\% accuracy in object identification but struggles with organization and localization.
However, these approaches cannot address complicated VR interactions. 
%and are not generic for different development ecosystems. 

%\end{itemize}

\vspace{-0.15cm}
\section{Conclusion}
\label{sec:Conclusion}
In this paper, we design the \textsc{EAT} framework, a generic three-layer abstraction framework based on OOP for modeling complex interaction behaviors and tasks in testing VR applications. 
Based on EAT, we present \textit{VRExplorer}, a novel model-based testing tool to achieve effective interactions with diverse virtual objects and explorations of complex VR scenes. 
To validate the performance of our approach, we evaluate \textit{VRExplorer} on \hn{11} representative VR projects. The experimental results validate our approach's superior performance to the SOTA method in terms of coverage and efficiency, as well as the ability to detect complicated real-world bugs. 

\section{Acknowledgement}
{\small 
This work was supported in part by the National Key Research and Development Program of China under Grant 2023YFB2704100, the National Natural Science Foundation of China (No. 62032025), the Seed Funding for Collaborative Research Grants of HKBU (with Grant No. RC-SFCRG/23-24/R2/SCI/06), the Major Key Project of Peng Cheng Laboratory under Grant PCL2025AS07.
}

\bibliographystyle{IEEEtran}
\bibliography{reference}

\end{document}